\documentclass[journal,twoside]{IEEEtran}
\usepackage[cmex10]{amsmath}
\usepackage{amsfonts,amssymb,url}

\newcommand{\bldc}{\boldsymbol{c}}
\newcommand{\bldh}{\boldsymbol{h}}
\newcommand{\bldp}{\boldsymbol{p}}
\newcommand{\bldq}{\boldsymbol{q}}
\newcommand{\bldr}{\boldsymbol{r}}
\newcommand{\bldu}{\boldsymbol{u}}
\newcommand{\bldv}{\boldsymbol{v}}
\newcommand{\bldx}{\boldsymbol{x}}
\newcommand{\bldy}{\boldsymbol{y}}
\newcommand{\bldgamma}{\boldsymbol{\gamma}}
\newcommand{\bldzero}{\boldsymbol{0}}
\newcommand{\bldone}{\boldsymbol{1}}
\newcommand{\bldH}{\boldsymbol{H}}
\newcommand{\bldI}{\boldsymbol{I}}
\newcommand{\bldL}{\boldsymbol{L}}

\newcommand{\cC}{\mathcal{C}}
\newcommand{\cI}{\mathcal{I}}
\newcommand{\cJ}{\mathcal{J}}
\newcommand{\cK}{\mathcal{K}}
\newcommand{\cM}{\mathcal{M}}
\newcommand{\cP}{\mathcal{P}}
\newcommand{\cS}{\mathcal{S}}
\newcommand{\cV}{\mathcal{V}}

\newcommand{\bH}{\bldH}
\newcommand{\bI}{\bldI}
\newcommand{\bL}{\bldL}
\newcommand{\weight}{\mathsf{w}}
\newcommand{\entropy}{\mathsf{H}}

\newcommand{\ff}{\mathbb{F}}
\newcommand{\nn}{\mathbb{N}}
\newcommand{\rr}{\mathbb{R}}

\newcommand{\conv}{\operatorname{conv}}
\newcommand{\GL}{\operatorname{GL}}
\DeclareMathOperator*{\argmin}{arg\,min} 

\newcommand{\define}{\stackrel{\mbox{\tiny $\triangle$}}{=}}
\newcommand{\AWGNC}{\mbox{\tiny AWGNC}}
\newcommand{\BSC}{\mbox{\tiny BSC}}
\newcommand{\BEC}{\mbox{\tiny BEC}}
\newcommand{\maxfrac}{\mbox{\tiny max-frac}}
\newcommand{\dual}{^{\perp}}

\newtheorem{theorem}{Theorem}[section]
\newtheorem{proposition}[theorem]{Proposition}
\newtheorem{lemma}[theorem]{Lemma}
\newtheorem{remark}[theorem]{Remark}
\newtheorem{corollary}[theorem]{Corollary}
\newtheorem{example}[theorem]{Example}
\newtheorem{definition}[theorem]{Definition}
\newtheorem{algorithm}[theorem]{Algorithm}

\newcommand{\mat}[1]{\left[\ \ \begin{matrix}#1\end{matrix}\;\ \right]}
\newcommand{\zo}{\mbox{\footnotesize \!$0$\!\! \normalsize}}
\newcommand{\ze}{\mbox{\footnotesize \!$\mathbf{1}$\!\! \normalsize}}

\begin{document}

\title{On the Pseudocodeword Redundancy\\ of Binary Linear Codes}

\author{Jens~Zumbr\"agel,~\IEEEmembership{Member,~IEEE,} Vitaly
  Skachek,~\IEEEmembership{Member,~IEEE,} and
  Mark~F.~Flanagan,~\IEEEmembership{Senior~Member,~IEEE}%
  \thanks{This work was supported in part by Science Foundation
    Ireland (Grants 06/MI/006, 08/IN.1/I1950, and 07/SK/I1252a).  The
    work of V.~Skachek was also supported in part by the National
    Research Foundation of Singapore (Research Grant
    NRF-CRP2-2007-03). The material in this paper was presented in
    part at the IEEE International Symposium on Information Theory,
    Austin, TX, USA, June 2010, and at the 19th International
    Symposium on Mathematical Theory of Networks and Systems,
    Budapest, Hungary, July 2010.}%
  \thanks{J.~Zumbr\"agel and M.~F.~Flanagan are with the Claude
    Shannon Institute, University College Dublin, Belfield, Dublin 4,
    Ireland.  E-mails: jens.zumbragel@ucd.ie,
    mark.flanagan@ieee.org.}%
  \thanks{V.~Skachek is with the Coordinated Science Laboratory,
    University of Illinois at Urbana-Champaign, 1308 W. Main Street,
    Urbana, IL 61801, USA.  E-mail: vitalys@illinois.edu. The work of
    this author was done in part while he was with the Claude Shannon
    Institute, University College Dublin, Belfield, Dublin 4, Ireland,
    and with the Division of Mathematical Sciences, School of Physical
    and Mathematical Sciences, Nanyang Technological University,
    21~Nanyang Link, 637371 Singapore.}%
}

\markboth{Submitted to IEEE Transactions on Information Theory}%
{Zumbr\"agel, Skachek, and Flanagan: On the Pseudocodeword Redundancy}

\maketitle


\begin{abstract}
  The AWGNC, BSC, and max-fractional pseudocodeword redundancies of a
  binary linear code are defined to be the smallest number of rows in
  a parity-check matrix such that the corresponding minimum
  pseudoweight is equal to the minimum Hamming distance of the code.
  It is shown that most codes do not have a finite pseudocodeword
  redundancy.  Also, upper bounds on the pseudocodeword redundancy for
  some families of codes, including codes based on designs, are
  provided.  The pseudocodeword redundancies for all codes of small
  length (at most $9$) are computed.  Furthermore, comprehensive
  results are provided on the cases of cyclic codes of length at most
  $250$ for which the eigenvalue bound of Vontobel and Koetter is
  sharp.
\end{abstract}



\section{Introduction}

\IEEEPARstart{P}{seudocodewords} represent the intrinsic mechanism of
failure of binary linear codes under linear-programming (LP) or
message-passing (MP) decoding (see, e.g., \cite{KV-characterization,
  KV-long-paper}).  The concept of \emph{pseudoweight} of a
pseudocodeword was introduced in~\cite{Wiberg} and~\cite{FKKR} (see
also~\cite{KV-long-paper}) as an analog to the pertinent parameter in
the maximum likelihood (ML) decoding scenario, i.e., the signal
Euclidean distance in the case of the additive white Gaussian noise
channel (AWGNC), or the Hamming distance in the case of the binary
symmetric channel (BSC).  Accordingly, for a binary linear code $\cC$
and a parity-check matrix $\bldH$ of $\cC$, the (AWGNC or BSC) minimum
pseudoweight $\weight_{\min}(\bldH)$ may be considered as a
first-order measure of decoder error-correcting performance for LP or
MP decoding.  Another closely related measure is the max-fractional
weight, which we sometimes also call pseudoweight in order to simplify
statements; it serves as a lower bound on both AWGNC and BSC
pseudoweights.

In order to minimize the decoding error probability under LP (or MP)
decoding, one might want to select a matrix~$\bH$ which maximizes the
minimum pseudoweight of the code for the given channel. Adding
redundant rows to the parity-check matrix introduces additional
constraints on the so-called \emph{fundamental cone}, and thus may
improve the performance of LP decoding and increase the minimum
pseudoweight.\footnote{We note that for message-passing iterative
  decoding, apart from the case of decoding over the binary erasure
  channel there is no general statement that additional parity-checks
  are beneficial.}  However, such additions increase the decoding
complexity under MP decoding, especially since linear combinations of
low-density rows may not yield a low-density result.  On the other
hand, there exist classes of codes for which sparse parity-check
matrices exist with many redundant rows, e.g.,
\cite{Kou_Lin_Fossorier}.

For the AWGNC, BEC (binary erasure channel), BSC, and max-fractional
pseudoweights, define $\rho_{\AWGNC}(\cC)$, $\rho_{\BEC}(\cC)$,
$\rho_{\BSC}(\cC)$, and $\rho_{\maxfrac}(\cC)$, respectively, to be
the minimum number of rows in any parity-check matrix $\bH$ such that
the minimum pseudoweight of $\cC$ with respect to this matrix is equal
to the code's minimum Hamming distance $d$.  For the sake of
simplicity, we sometimes use the notation $\rho(\cC)$ when the type of
channel is clear from the context.  The value $\rho(\cC)$ is called
the (AWGNC, BEC, BSC, or max-fractional) \emph{pseudocodeword
  redundancy} (or pseudoredundancy) of $\cC$. If for the code $\cC$
there exists no such matrix $\bH$, we say that the pseudoredundancy is
infinite.

The BEC pseudocodeword redundancy, which is equivalent to the
\emph{stopping redundancy}, is studied in~\cite{Schwartz_Vardy}, where
it is shown that for any linear code the BEC pseudoredundancy is
finite; the paper also contains bounds on $\rho_{\BEC}(\cC)$ for
general binary linear codes, and for some specific families of codes.
These bounds were subsequently improved, for instance
in~\cite{Han_Siegel}.  The study of BSC pseudoredundancy was initiated
in~\cite{Kelley_Sridhara}, where the authors presented bounds on
$\rho_{\BSC}(\cC)$ for various families of codes.

In this work, we further investigate pseudoredundancy for the AWGNC,
BSC, and max-fractional pseudoweight. We show that for most codes
there exists no $\bH$ such that the minimum pseudoweight (with respect
to $\bH$) is equal to $d$, and therefore the AWGNC, BSC, and
max-fractional pseudocodeword redundancy (as defined above) is
infinite for most codes.  For some code families for which the
pseudoredundancy is finite, we provide upper bounds on its value.  We
consider in particular constructions of new codes from old and codes
based on designs.  Furthermore, we compute the pseudocodeword
redundancies for all codes of small length (at most $9$), and we
investigate cyclic codes for which the eigenvalue bound of Vontobel
and Koetter~\cite{KV-lower-bounds} is sharp.

The outline of the paper is as follows.  In Section
\ref{sec:general_settings} we provide detailed definitions and some
background information on LP decoding, pseudocodewords, the minimum
pseudoweight, and the pseudocodeword redundancy; we also discuss
related notions appearing in the literature.  Subsequently, we show in
Section~\ref{sec:random_codes} that the pseudocodeword redundancy for
random codes is infinite with high probability.  The next four
sections are concerned with upper bounds on the pseudoredundancy for
some particular classes of codes; we investigate punctured codes and
codes of dimension 2 in Section~\ref{sec:puncture}, constructions of
codes from other codes in Section~\ref{sec:inheritance}, parity-check
matrices of row-weight 2 in Section~\ref{sec:row_weight_2}, and codes
based on designs in Section~\ref{sec:designs}.  The final two sections
are devoted to experimental results; Section~\ref{sec:short_codes}
examines the pseudocodeword redundancy for all codes of small length,
and Section~\ref{sec:kv_bound} deals with cyclic codes that meet the
eigenvalue bound on the minimum AWGNC pseudoweight by Vontobel and
Koetter.


\section{General Settings}\label{sec:general_settings}

Let $\ff_2$ be the binary field and let $\rr$ be the field of real
numbers.  Addition and multiplication (including matrix-vector and
matrix-matrix multiplication) are carried out in $\ff_2$ when the
operands are defined over $\ff_2$, and in $\rr$ when the operands are
defined over the reals.  Occasionally, we will explicitly convert
elements in $\ff_2$ into real numbers; in this case we identify
$0\in\ff_2$ with $0\in\rr$ and $1\in\ff_2$ with $1\in\rr$.

Let $\cC$ be a code of length $n \in \nn$ over the binary field
$\ff_2$, defined by \[ \cC = \ker \bH = \{ \bldc \in \ff_2^n \mid \bH
\bldc^T = \bldzero^T \} \] where $\bH$ is an $m \times n$
\emph{parity-check matrix} over $\ff_2$ of the code~$\cC$. Obviously,
the code $\cC$ may admit more than one parity-check matrix, and all
the codewords form a linear vector space of dimension $k \ge n-m$.  We
say that $k$ is the \emph{dimension} of the code $\cC$.  We denote by
$d(\cC)$ (or just $d$) the minimum Hamming distance (also called the
minimum distance) of $\cC$. The code $\cC$ may then be referred to as
an $[n,k,d]$ linear code over $\ff_2$.

Denote the set of column indices and the set of row indices of the
parity-check matrix $\bH$ by $\cI = \{1,\dots,n\}$ and $\cJ =
\{1,\dots,m\}$, respectively.  For any row index $j\in\cJ$ we let
$\cI_j \define \{ i \in \cI \mid H_{j,i} \neq 0 \}$ denote the set of
the column indices where the parity-check matrix is nonzero; similarly
for any column index $i\in\cI$ we let $\cJ_i \define \{ j \in \cJ \mid
H_{j,i} \neq 0 \}$ denote the corresponding set of row indices.

The matrix $\bH$ is said to be \emph{$(w_c,w_r)$-regular} if
$|\cJ_i|=w_c$ for all $i\in\cI$ and $|\cI_j|=w_r$ for all $j\in\cJ$; a
$(w,w)$-regular matrix is also called simply \emph{$w$-regular}.

\subsection{LP decoding}

We give a brief review of LP decoding.  Consider data transmission
over a memoryless binary-input output-symmetric channel with channel
law $p_{Y|X}(y|x)$.  Based on the received vector $\bldy =
(y_1,\dots,y_n)$ we can define the log-likelihood-ratio vector
$\bldgamma = (\gamma_1,\dots,\gamma_n)\in\rr^n$ by $\gamma_i \define
\log(p_{Y|X}(y_i|0))-\log(p_{Y|X}(y_i|1))$ for $i\in\cI$.  Viewing the
code $\cC$ canonically as a subset of $\{0,1\}^n\subset \rr^n$, one
can then express ML decoding as the minimization problem
\[ \hat{\bldx} \define \argmin_{\bldx\in\cC}\, \langle \bldx, \bldgamma
\rangle \:. \]
This is equivalent to the linear programming problem
\[ \hat{\bldx} \define \argmin_{\bldx\in\conv(\cC)} \langle \bldx,
\bldgamma \rangle \:, \] where $\conv(\cC)$ denotes the convex hull
of $\cC$ in $\rr^n$.  However, since the number of defining
hyperplanes of $\conv(\cC)$ usually grows exponentially with the block
length, this minimization problem becomes impractical.

Instead one might consider a relaxation of the above minimization
problem (see \cite{Feldman, Feldman-et-al, KV-long-paper}), where the
convex hull $\conv(\cC)$ is replaced by the so-called fundamental
polytope $\cP(\bH)$ to be defined next.  For $j\in \cJ$, let $\bldh_j$
denote the $j$-th row of the parity-check matrix $\bH$, and consider
the local code
\[  \cC_j  = \{ \bldc \in \ff_2^n \mid \bldh_j \bldc^T = 0 \} \]
consisting of all binary vectors satisfying the $j$-th parity-check,
so that $\cC = \bigcap_{j\in\cJ}\cC_j$.
Then the \emph{fundamental polytope} $\cP \define \cP(\bH)$ is defined as
\[ \cP \define \bigcap_{j\in\cJ} \conv(\cC_j) \:, \] 
where again $\cC_j$ is viewed as a subset of $\rr^n$.  Now \emph{LP
  decoding} of a binary linear code $\cC$ with parity-check matrix
$\bH$ can be expressed as the minimization problem
\begin{equation}\label{eq:lp-decoding}
  \hat{\bldx} \define \argmin_{\bldx\in\cP}\, \langle \bldx,
  \bldgamma \rangle \:,
\end{equation}
where $\cP = \cP(\bH)$ denotes the fundamental polytope.

We note that $\conv(\cC)\subseteq\cP$, where the inclusion is usually
proper.  However, the number of defining hyperplanes of $\cP$ is
typically much smaller than for $\conv(\cC)$, in particular for LDPC
codes, so that the corresponding linear programming problem becomes
tractable.

If $\cP$ is strictly larger than $\conv(\cC)$ then it may happen that
the decoding rule~(\ref{eq:lp-decoding}) outputs a vertex\footnote{The
  set of optimal solutions contains a vertex, and one may assume that
  the output is a vertex.}  of $\cP$ that is not a vertex of
$\conv(\cC)$, i.e., not a codeword.  Such vertices, called
pseudocodewords, are the reason for the suboptimality of LP decoding
with respect to ML decoding.

Note that the fundamental polytope $\cP(\bH)$ is dependent on the
parity-check matrix $\bH$ rather than the code $\cC$ itself, but we
always have $\cP(\bH)\cap \{0,1\}^n = \cC$, cf.\ \cite{Feldman,
  Feldman-et-al}.

\subsection{The fundamental cone and pseudoweights}

When analyzing LP decoding, we may assume without loss of generality
that the zero codeword $\bldzero$ has been sent; then, given this
assumption, the probability of correct LP decoding depends only on the
conic hull of the fundamental polytope rather than on the fundamental
polytope itself (see \cite{Feldman, Feldman-et-al, KV-long-paper}).
The conic hull of the fundamental cone $\cP(\bH)$ is called the
\emph{fundamental cone} $\cK(\bH)$.  More concretely, $\cK(\bH)$ is
given as the set of vectors $\bldx\in \rr^n$ that satisfy
\begin{equation}\label{eq:fundcone-inequality-1}
  \forall j \in \cJ, \; \forall \ell \in \cI_j  :\ \:
  x_\ell \le \sum_{i \in \cI_j \setminus \{ \ell \}} x_i \; ,
\end{equation}
\begin{equation}\label{eq:fundcone-inequality-2}
  \forall i \in \cI :\ \: x_i \ge 0 \; .
\end{equation}

The vectors $\bldx\in\cK(\bH)$ are called \emph{pseudocodewords}%
\footnote{Some authors consider only the vertices of the fundamental
  polytope $\cP(\bH)$ as pseudocodewords, but we will use this more
  general definition which includes all vectors of the fundamental
  cone $\cK(\bH)$.} of~$\cC$ with respect to the parity-check matrix
$\bH$.  Note again that the fundamental cone $\cK(\bH)$ depends on the
parity-check matrix $\bH$ rather than on the code $\cC$ itself.  At
the same time, the fundamental cone is independent of the underlying
communication channel.

\begin{example}\label{exa:fund_cone}
  Let $\cC$ be the $[7,4,3]$ Hamming code with parity-check matrix
  \[ \bH = \mat{ 
    \ze & \ze & \ze & \zo & \ze & \zo & \zo \\
    \zo & \ze & \ze & \ze & \zo & \ze & \zo \\
    \zo & \zo & \ze & \ze & \ze & \zo & \ze } . \]
  Then the fundamental cone inequalities read:
  {\small
  \begin{align*}
    & x_1 \le x_2+x_3+x_5 \quad x_2\le x_3+x_4+x_6 \quad x_3\le x_4+x_5+x_7 \\
    & x_2 \le x_1+x_3+x_5 \quad x_3\le x_2+x_4+x_6 \quad x_4\le x_3+x_5+x_7 \\
    & x_3 \le x_1+x_2+x_5 \quad x_4\le x_2+x_3+x_6 \quad x_5\le x_3+x_4+x_7 \\
    & x_5 \le x_1+x_2+x_3 \quad x_6\le x_2+x_3+x_4 \quad x_7\le x_3+x_4+x_5 \\
    & 0 \le x_1 \quad 0 \le x_2 \quad 0 \le x_3 \quad 0 \le x_4 
    \quad 0 \le x_5 \quad 0 \le x_6 \quad 0 \le x_7 
  \end{align*}}
\end{example}

The influence of a nonzero pseudocodeword on the decoding performance
will be measured by its \emph{pseudoweight}, which depends on the
channel at hand.  The BEC, AWGNC, BSC pseudoweights, and max-fractional
weight of a nonzero pseudocodeword $\bldx \in \cK(\bH)$ were defined
in~\cite{FKKR} and~\cite{KV-long-paper} as follows:
\begin{align*}
  \weight_{\BEC} (\bldx) & \,\define\, 
  \left| \mbox{supp} ( \bldx ) \right| \; , \\
  \weight_{\AWGNC} (\bldx) & \,\define\,  
  \frac{\left( \sum_{i \in \cI} x_i \right)^2}{\sum_{i \in \cI} x_i^2} \: .
\end{align*}
Let $\bldx'$ be a vector in $\rr^n$ with the same components as
$\bldx$ but in non-increasing order.  For $i-1 < \xi \le i$, where $1
\le i \le n$, let $\phi(\xi) \stackrel{\triangle}{=} x'_i$. Define
$\Phi(\xi) \define \int_{0}^{\xi} \phi(\xi') \; d \xi'$ and \[
\weight_{\BSC}(\bldx) \define 2\, \Phi^{-1} ( \Phi(n)/2 ) \; . \] 
Finally, the max-fractional weight of $\bldx$ is defined as
\[\weight_{\maxfrac} (\bldx) \,\define\,
\frac{\sum_{i \in \cI} x_i}{\max_{i \in \cI} x_i} \: .\]

Additionally, the pseudoweight of the all-zero vector is usually
defined to be zero, i.e., $\weight(\bldzero) = 0$, for all four
pseudoweights $\weight$, but this is inessential for this paper.

Note that for binary vectors $\bldx\in\{0,1\}^n\setminus\{\bldzero\}$
we have
\[ \weight_{\BSC}(\bldx) = \weight_{\AWGNC}(\bldx) =
\weight_{\BSC}(\bldx) = \weight_{\maxfrac}(\bldx) =
\weight_{\mathrm{H}}(\bldx) \:, \] where $\weight_{\mathrm{H}}(\bldx)$
denotes the Hamming weight of $\bldx$.

\begin{example}
  Let $\cC$ and $\bH$ be as in Example~\ref{exa:fund_cone}.  The
  vector $\bldx = (0,0,1,0,1,1,2)$ is a pseudocodeword in $\cK(\bH)$
  with weights $\weight_{\BEC} (\bldx) = \left| \mbox{supp}
    (\bldx) \right| = 4$ and $\weight_{\AWGNC} (\bldx) = \left(
    \sum_{i \in \cI} x_i \right)^2 / \sum_{i \in \cI} x_i^2 = 25/7$.
  Furthermore, $\weight_{\BSC}(\bldx) =
  2\,\Phi^{-1}\big((\sum_{i\in\cI}x_i)/2\big) = 2\,\Phi^{-1}(5/2) =
  3$, where $\bldx'= (2,1,1,1,0,0,0)$, and finally,
  $\weight_{\maxfrac} (\bldx) = \sum_{i \in \cI} x_i / \max_{i \in
    \cI} x_i = 5/2$.
\end{example}

We define the BEC \emph{minimum pseudoweight} of the code $\cC$ with
respect to the parity-check matrix $\bH$ as
\[ \weight_{\min}^{\BEC} (\bH) \,\define\, \min_{\bldx\in\cK(\bH)
  \setminus \{ \bldzero \} } \weight_{\BEC} (\bldx) \; .\] The
quantities $\weight_{\min}^{\AWGNC} (\bH) $, $\weight_{\min}^{\BSC}
(\bH) $ and $\weight_{\min}^{\maxfrac} (\bH)$ are defined similarly.
We note that the considered pseudoweights are invariant under scaling
by a positive scalar, and that a minimum is indeed attained on
$\cK(\bH)\setminus\{0\}$ (see \cite[Sect.~6]{KV-long-paper}).  When
the type of pseudoweight is clear from the context, we sometimes use
the notation $\weight_{\min} (\bH)$.  Note that all four minimum
pseudoweights are upper bounded by~$d$, the code's minimum distance.

\subsection{Pseudocodeword redundancy}

Given a code $\cC$ we will define the pseudocodeword redundancy as the
minimum number of rows in a parity-check matrix $\bH$ for $\cC$ such
that the corresponding minimum pseudoweight equals the minimum
distance.

So for a binary linear $[n,k,d]$ code $\cC$ we define the BEC
\emph{pseudocodeword redundancy} of the code $\cC$ as \[
\rho_{\BEC}(\cC) \,\define\, \inf\{\#\text{rows}(\bH) \mid
\ker\bH=\cC\,,\, \weight_{\min}^{\BEC}(\bH)=d\} \: ,\] where
$\inf\varnothing\define\infty$, and similarly we define the
pseudocodeword redundancies $\rho_{\AWGNC}(\cC)$, $\rho_{\BSC}(\cC)$,
and $\rho_{\maxfrac}(\cC)$ for the AWGNC and BSC pseudoweights, and
the max-fractional weight.  When the type of pseudocodeword redundancy
is clear from the context, we sometimes use the notation $\rho(\cC)$.

We remark that all pseudocodeword redundancies satisfy $\rho(\cC) \ge
r\define n-k$.

\begin{example}
  Let $\cC$ be the $[7,4,3]$ Hamming code. Then:
  \begin{gather*}
    \rho_{\maxfrac}(\cC) = 7 \; \ge \;
    \rho_{\AWGNC}(\cC) = 3 \; \ge \;
    \rho_{\BEC}(\cC) = 3 \\
    \rho_{\maxfrac}(\cC) = 7 \; \ge \;
    \rho_{\BSC}(\cC) = 4 \; \ge \;
    \rho_{\BEC}(\cC) = 3
  \end{gather*}
  The following matrices $\bH_3$, $\bH_4$, and $\bH_7$ are examples
  for parity-check matrices with a minimum number of rows such that
  $\weight_{\min}^{\BEC}(\bH_3) = \weight_{\min}^{\AWGNC}(\bH_3) = 3$,
  $\weight_{\min}^{\BSC}(\bH_4) = 3$, and
  $\weight_{\min}^{\maxfrac}(\bH_7) = 3$ holds.
  \begin{gather*}
    \bH_3 = \mat{ 
      \ze & \ze & \ze & \zo & \ze & \zo & \zo \\
      \zo & \ze & \ze & \ze & \zo & \ze & \zo \\
      \zo & \zo & \ze & \ze & \ze & \zo & \ze } \\[1ex]
    \bH_4 = \mat{
      \ze & \ze & \zo & \ze & \zo & \zo & \ze \\
      \ze & \zo & \ze & \zo & \ze & \zo & \ze \\
      \zo & \ze & \ze & \zo & \zo & \ze & \ze \\
      \zo & \zo & \zo & \ze & \ze & \ze & \ze } \\[1ex]
    \bH_7 = \mat{
      \ze & \ze & \ze & \zo & \ze & \zo & \zo \\
      \zo & \ze & \ze & \ze & \zo & \ze & \zo \\
      \zo & \zo & \ze & \ze & \ze & \zo & \ze \\
      \ze & \zo & \zo & \ze & \ze & \ze & \zo \\
      \zo & \ze & \zo & \zo & \ze & \ze & \ze \\
      \ze & \zo & \ze & \zo & \zo & \ze & \ze \\
      \ze & \ze & \zo & \ze & \zo & \zo & \ze } 
  \end{gather*}
  These matrices were found by computer search, see
  Section~\ref{sec:short_codes}.
\end{example}

We describe the behavior of the pseudocodeword redundancy and the
minimum pseudoweight for a given binary linear $[n,k,d]$ code~$\cC$
by introducing four classes of codes:\smallskip

\begin{description}[\IEEEsetlabelwidth{{\bf (class 0)}}]
\item[{\bf (class 0)}] $\rho(\cC)$ is infinite, i.e., there is no
  parity-check matrix~$\bH$ with $d=\weight_{\min}(\bH)$,
\item[{\bf (class 1)}] $\rho(\cC)$ is finite, but $\rho(\cC)>r$,
\item[{\bf (class 2)}] $\rho(\cC)=r$, but $\cC$ is not in
  class 3,
\item[{\bf (class 3)}] $d=\weight_{\min}(\bH)$ for \emph{every}
  parity-check matrix $\bH$ of~$\cC$.
\end{description}

Note that if a code has infinite pseudocodeword redundancy, then LP
decoding for this code can never achieve the ML decoding performance;
on the other hand, if a code's pseudocodeword redundancy is finite,
its value gives a (very approximate) indication of the LP decoding
complexity required to achieve this bound. Note that this is a
fundamental complexity associated with the code, and not tied to a
particular parity-check matrix. We leave it as a direction for further
research to provide more general definitions which capture the average
complexity-performance tradeoff of LP decoding as more redundant rows
are added to the parity-check matrix.

\subsection{Basic Connections} 

The different minimum pseudoweights are related as follows.  This
result is taken from~\cite{KV-long-paper}.

\begin{lemma}\label{lemma:relations}
  Let $\cC$ be a binary linear code with the parity-check matrix
  $\bH$.  Then,
  \begin{gather*}
    \weight_{\min}^{\maxfrac} (\bH) \; \le \; 
    \weight_{\min}^{\AWGNC} (\bH) \; \le \; 
    \weight_{\min}^{\BEC} (\bH) \; , \\
    \weight_{\min}^{\maxfrac} (\bH) \; \le \; 
    \weight_{\min}^{\BSC} (\bH) \; \le \; 
    \weight_{\min}^{\BEC} (\bH) \; . 
  \end{gather*}
\end{lemma}

As a straightforward corollary we obtain the following theorem, which
relates the different pseudoredundancies.

\begin{theorem}\label{thm:pseudoredundancies}
  Let $\cC$ be a binary linear code.  Then,
  \begin{gather*}
    \rho_{\maxfrac} (\cC) \; \ge \; \rho_{\AWGNC} (\cC) \; \ge \; 
    \rho_{\BEC} (\cC) \; , \\
    \rho_{\maxfrac} (\cC) \; \ge \; \rho_{\BSC} (\cC) \; \ge \; 
    \rho_{\BEC} (\cC) \; .
  \end{gather*}
\end{theorem}

\subsection{Related Notions}\label{sec:related}

As mentioned in the introduction, Schwartz and Vardy consider
in~\cite{Schwartz_Vardy} the so-called stopping distance of a binary
linear code given by a parity-check matrix, and the stopping
redundancy of a binary linear code.  With \cite[Proposition
51]{KV-long-paper} it is easy to see that the stopping distance equals
the minimum BEC pseudoweight, and thus the stopping redundancy is
equivalent to the BEC pseudocodeword redundancy.

Besides pseudocodewords, the notion of \emph{trapping set}
\cite{trapping} is another concept for analyzing the performance of
binary linear codes under MP decoding.  In \cite{LHMH-trapping} the
{\em trapping redundancy} for binary linear codes is introduced as a
generalization of the stopping redundancy, and several upper bounds
are presented.\smallskip

In~\cite{Kashyap} a binary linear code $\cC$ is called
\emph{geometrically perfect} if it admits a parity-check matrix $\bH$
such that the fundamental polytope equals the convex hull of the code,
i.e., $\cP(\bH) = \conv(\cC)$.
In this case ML decoding can be exactly described as an instance of LP
decoding.  Kashyap~\cite[Theorem~VI.2]{Kashyap} gave a
characterization of all geometrically perfect codes: a binary linear
code $\cC$ is geometrically perfect if and only if $\cC$ does not
contain as a minor\footnote{A \emph{minor} of a code ${\cC}$ is any
  code obtained from ${\cC}$ by a (possibly empty) sequence of
  shortening and puncturing operations.}  any code equivalent to
certain codes $\cC_1$, $\cC_2$, $\cC_3$ with parameters $[7,3,4]$,
$[10,5,4]$, and $[10,4,4]$, respectively.

It is easy to see that for geometrically perfect codes all four
pseudocodeword redundancies are finite.\smallskip

Smarandache and Vontobel~\cite{Smarandache_Vontobel} define the
\emph{pseudoweight spectrum gap} for a binary linear code $\cC$ given
by a parity-check matrix $\bH$ as follows.  The set $\cM(\bH)$ of all
\emph{minimal pseudocodewords} is defined as the set of all vectors
$\bldx\in\rr^n$ that lie on an edge of the fundamental cone
$\cK(\bH)$.  Now let $\cM'(\bH)$ denote the set of all minimal
pseudocodewords that are not scalar multiples of codewords
$\bldc\in\cC$, and let $\weight$ be any of the BEC, AWGNC, BSC, or
max-fractional pseudoweight.  Then the pseudoweight spectrum gap is
the quantity \[ g(\bH) \define \min_{\bldx\in\cM'(\bH)} \weight(\bldx)
- d(\cC) \:. \] It is apparent that $g(\bH)\ge \weight_{\min}(\bH) -
d(\cC)$, and we have $\weight_{\min}(\bH) = d(\cC)$ if and only if
$g(\bH)\ge 0$.

If the pseudoweight spectrum gap $g(\bH)$ is strictly positive then
the LP decoding performance approaches ML decoding performance as the
signal-to-noise ratio goes to infinity.  To date, only few examples of
interesting codes with positive pseudoweight spectrum gap are known;
these include the codes based on the Euclidean plane or the projective
plane \cite[Theorem~8]{Smarandache_Vontobel}.


\section{Pseudoredundancy of Random Codes}\label{sec:random_codes} 

In this section we show that for most binary linear codes the AWGNC
and BSC pseudoredundancies are infinite.  We begin with the following
lemma.

\begin{lemma}\label{lemma:awgn}
  For a binary linear code $\cC$ of length $n$, let $d\dual$ be
  the minimum distance of the dual code.  Then, the minimum AWGNC
  pseudoweight of $\cC$ (with respect to any parity-check matrix
  $\bH$) satisfies
  \begin{equation}\label{eq:pseudo-distance-bound-random}
    \weight_{\min}^{\AWGNC} \le 
    \frac{(n + d\dual - 2)^2}{(d\dual-1)^2 + (n-1)} \; . 
  \end{equation}
\end{lemma}

\begin{IEEEproof} 
  Consider the pseudocodeword $\bldx = (x_1, x_2, \dots, x_n) \define
  (d\dual \!\!-\! 1 \,,\, 1 \,,\, \dots, 1 )$.  Since $d\dual$ is the
  minimum distance of the dual code, every row in $\bH$ has weight at
  least $d\dual$.  Therefore, all
  inequalities~(\ref{eq:fundcone-inequality-1})
  and~(\ref{eq:fundcone-inequality-2}) are satisfied for this $\bldx$,
  and so it is indeed a legal pseudocodeword. Finally, observe that
  the AWGNC pseudoweight of $\bldx$ is given by the right-hand side of
  (\ref{eq:pseudo-distance-bound-random}).
\end{IEEEproof}

In the sequel, we use the term \emph{random code} for a binary linear
code $\cC$ whose $k \times n$ generator matrix contains independently
and uniformly distributed random entries from $\ff_2$. The following
result is known as the Gilbert-Varshamov bound.  If we pick a code by
selecting the generator matrix entries at random, the resulting code
$\cC$ has rate $R = k/n$ and relative minimum distance $\delta$, such
that
\[ \delta \ge \entropy_2^{-1} (1 - R) - \epsilon \; , \] with
probability approaching $1$ as $n \rightarrow \infty$, for any fixed
small $\epsilon > 0$, where $\entropy^{-1}_2(\cdot)$ is the inverse of
the binary entropy function $\entropy_2 (p) = - p \log_2 p - (1 - p)
\log_2 (1 - p)$ for $p\in[0\,,\, 1/2]$.  A similar result also holds
when the code $\cC$ is defined by selecting the parity-check matrix
entries (independently and uniformly) at random.

Let $R = k/n$ be fixed.  Then, if we select at random a $k \times n$
matrix over $\ff_2$, which corresponds to a code $\cC$, the relative
minimum distance of $\cC$ is at least $\entropy_2^{-1} (1 - R) -
\epsilon$ (with probability approaching~$1$ as $n \rightarrow \infty$)
and the relative minimum distance of the dual code of $\cC$ is at
least $\entropy_2^{-1} (R) - \epsilon$ (again, with probability
approaching~$1$ as $n \rightarrow \infty$).  By taking the
intersection of these two events, both the code and the dual code have
relative minimum distances which are $\epsilon$-close to the
Gilbert-Varshamov bound with probability approaching~$1$ as $n
\rightarrow \infty$.  (The reader can refer to~\cite[Theorems 4.4,
4.5, and 4.10]{Roth-book} and to~\cite[Theorem 8 and Exercise
3]{Guruswami-notes}.)

To this end, we take a random binary linear code $\cC$ 
of arbitrary length $n$ (for $n \rightarrow \infty$) with $R = k/n$. 
The dual code $\cC\dual$ of $\cC$, with probability close to one, has
rate $R\dual=1-R$ and relative minimum distance $\delta\dual = d\dual/n$ that
attains the Gilbert-Varshamov bound
\[ \delta\dual \ge \mu \define \entropy^{-1}_2 (1 - R\dual) - \epsilon
= \entropy^{-1}_2(R) - \epsilon \; , \] 

Note that~(\ref{eq:pseudo-distance-bound-random}) may be written in
terms of the relative minimum distance $\delta\dual$ of the dual code
as follows:
\begin{equation}
  \weight_{\min}^{\AWGNC} \le 
  \frac{(1 + \delta\dual\! - 2/n)^2}
  {(\delta\dual\!- 1/n)^2 + (1/n-1/n^2)} \; .
  \label{eq:pseudo-distance-new}
\end{equation}
Hence, for large $n$, the minimum pseudoweight of the code $\cC$ is
bounded from above by $(1 + 1/\delta\dual)^2 + \epsilon' \le
(1+1/\mu)^2 + \epsilon'$ for some small $\epsilon' > 0$, and this
bound does not depend on $n$. On the other hand, $\cC$ is a random
code and so its minimum distance satisfies the Gilbert-Varshamov
bound, namely
\[ d\ge \left( \entropy^{-1}_2 (1 - R) - \epsilon \right) \cdot n \;
,\] which increases linearly with $n$ for a fixed $R$. This
immediately establishes the following theorem.

\begin{theorem}\label{thm:non-existence_AWGNC}
  Let $0<R<1$ be fixed.  For a random binary linear code $\cC$ of
  length $n$ and rate~$R$, there is, with probability approaching~$1$
  as $n$ tends to infinity, a gap between the minimum AWGNC
  pseudoweight (with respect to any parity-check matrix) and the
  minimum distance.  Therefore, the AWGNC pseudoredundancy is infinite
  for most codes.
\end{theorem}

\begin{remark}
  The result in Theorem~\ref{thm:non-existence_AWGNC} is different
  from, but related to, the results in Propositions~49 and
  Corollary~50 in~\cite{KV-long-paper}, where it was shown that the
  minimum AWGN pseudoweight of ensembles of regular LDPC codes grows
  sublinearly in the code length.  Indeed, there are three fundamental
  differences between our results and~\cite{KV-long-paper}: (i) We do
  not assume anything about the density of the parity-check matrix
  $\bH$.  We also use the fact that the dual code of the random code
  is asymptotically good; for a regular LDPC code this is not
  true. (ii) We consider the fundamental cone, which is formed by all
  possible linear combinations of the rows of $\bH$; by contrast, the
  authors of~\cite{KV-long-paper} consider only the case when the
  column weight of $\bH$ is smaller than its row weight. (iii) We show
  that the minimum pseudoweight of the considered ensemble is bounded
  from above by a constant, while in~\cite{KV-long-paper} this
  quantity is shown to be bounded by a sublinear function.
\end{remark}\smallskip

The following lemma is a counterpart of Lemma~\ref{lemma:awgn} for the
BSC.

\begin{lemma}\label{lemma:bsc}
  Let $\cC$ be a binary linear code of length $n$, and let $d\dual$
  be the minimum distance of the dual code.  Then, the minimum BSC
  pseudoweight of $\cC$ (with respect to any parity-check matrix
  $\bH$) satisfies
  \[ \weight_{\min}^{\BSC} \le 2 \lceil n/{d\dual} \rceil \; . \]
\end{lemma}

\begin{IEEEproof}
  Consider the pseudocodeword 
  \[ \bldx = (x_1, x_2, \dots, x_n) \define (\underbrace{ 
    d\dual\!\!-\!1, \dots, d\dual\!\!-\!1}_\tau \,,\, 
  \underbrace{1, \dots, 1}_{n-\tau}) \; , \]
  for some positive integer $\tau$. This vector $\bldx$ is then a
  legal pseudocodeword; since $d\dual$ is the minimum distance of the
  dual code, every row in $\bH$ has a weight of at least $d\dual$, and
  so, all inequalities~(\ref{eq:fundcone-inequality-1})
  and~(\ref{eq:fundcone-inequality-2}) are satisfied by this $\bldx$.

  If $\tau(d\dual\!-1) \ge n - \tau$ then by the definition of the BSC
  pseudoweight $\weight_{\BSC} (\bldx) \le 2 \tau$. This condition is
  equivalent to $\tau d\dual \ge n$. Therefore, we set $\tau = \lceil
  n/{d\dual} \rceil$. For the corresponding vector $\bldx$, the
  pseudoweight is less or equal to $2\tau = 2 \lceil n/{d\dual}
  \rceil$.
\end{IEEEproof}

Similarly to the AWGNC case, let $\cC$ be a random binary linear code
of length $n$ with $R=k/n$.  The parameters $R\dual$
and~$\delta\dual$ of its dual code $\cC\dual$ attain with high
probability the Gilbert-Varshamov bound $\delta\dual\ge\mu$.

From Lemma~\ref{lemma:bsc}, for all $n$, the pseudoweight of the
code~$\cC$ is bounded from above by
\[2 \lceil n/{d\dual} \rceil < 2/\delta\dual + 2 \le 2/\mu + 2 \; ,\]
which is a constant.  On the other hand, $\cC$ is a random code and
its minimum distance also satisfies the Gilbert-Varshamov bound, so it
increases linearly with $n$. This proves the following theorem.

\begin{theorem}\label{thm:non-existence_BSC}
  Let $0<R<1$ be fixed.  For a random binary linear code $\cC$ of
  length $n$ and rate~$R$, there is, with probability approaching~$1$
  as $n$ tends to infinity, a gap between the minimum BSC pseudoweight
  (with respect to any parity-check matrix) and the minimum distance.
  Therefore, the BSC pseudoredundancy is infinite for most codes.
\end{theorem}

The last theorem disproves the conjecture in~\cite{Kelley_Sridhara}
that the BSC pseudoredundancy is finite for all binary linear
codes.\footnote{We note that a slightly different definition of BSC
  pseudoweight was given in~\cite{Kelley_Sridhara}, but the statement
  of Lemma~\ref{lemma:bsc} and thus
  Theorem~\ref{thm:non-existence_BSC} hold with the same proof also
  with respect to this definition.}

\begin{example}
  Consider the [23,12] Golay code having minimum distance $d=7$. The
  minimum distance of its dual code is $d\dual=8$.  We can take a
  pseudocodeword $\bldx$ as in the proof of Lemma~\ref{lemma:bsc} with
  $\tau = \lceil n/{d\dual} \rceil = 3$.  We have $\weight_{\BSC}
  (\bldx) \le 2 \tau = 6$, thus obtaining that the minimum distance is
  not equal to the minimum pseudoweight.

  Similarly, for the [24,12] extended Golay code we have $d=d\dual=8$,
  and by taking $\tau = \lceil n/{d\dual} \rceil = 3$ we obtain
  $\weight_{\BSC} (\bldx) \le 2\tau = 6$.

  Note however that the presented techniques do not answer the
  question of whether these Golay codes have finite AWGNC
  pseudoredundancy.
\end{example}

In the context of the extended Golay code we mention that there other
interesting graphical representations of codes than by Tanner graphs;
in particular, a minimal \emph{tail-biting trellis} has been
constructed for the extended Golay code in \cite{TBT}.  The
pseudoweights of its pseudocodewords are investigated in \cite{FKKR},
where it is shown that there are pseudocodewords with a BSC
pseudoweight of $6$; on the other hand, as far as we know, it is still
unknown whether there are nonzero pseudocodewords of the tail-biting
trellis with an AWGNC pseudoweight of less than $8$.

We have seen in this section that the AWGNC pseudoredundancy and the
BSC pseudoredundancy of a random binary linear code is infinite.  From
Theorem~\ref{thm:pseudoredundancies} it follows that this holds also
for the pseudoredundancy with respect to the max-fractional weight.


\section{Basic Upper Bounds}\label{sec:puncture}

Whereas a random code has infinite pseudoredundancy for the AWGNC and
the BSC, there are several families of codes for which the
pseudoredundancy is finite.  Sections~\ref{sec:puncture},
\ref{sec:inheritance}, \ref{sec:row_weight_2}, and~\ref{sec:designs}
deal with upper bounds on the pseudoredunancy for some particular
classes of codes.

We start with this section considering two basic situations, namely
the puncturing of zero coordinates and codes of minimum distance~$2$.
The following results hold with respect to the BEC, AWGNC, and BSC
pseudoweights, and the max-fractional weight.

\begin{lemma}\label{lemma:puncturing}
  Let $\cC$ be an $[n,k,d]$ code having $t$ zero coordinates, and let
  $\cC'$ be the $[n-t,k,d]$ code obtained by puncturing~$\cC$ at these
  coordinates.  Then 
  \[ \rho(\cC')\le \rho(\cC)\le \rho(\cC')+t \:. \]
\end{lemma}

\begin{IEEEproof}
  For notational purposes, we identify $\rr^n$ with $\rr^{\cI}$, and
  for $\bldx\in\rr^{\cI}$ and some subset $\cI'\subseteq\cI$ we let
  $\bldx|_{\cI'}\in\rr^{\cI'}$ be the projection of $\bldx$ onto the
  coordinates in $\cI'$.

  Let $\cI'\subseteq\cI$ be the set of nonzero coordinates of the
  code~$\cC$.  To prove the first inequality, let $\bH$ be a
  $\rho\times n$ parity-check matrix for $\cC$.  Consider its
  $\rho\times(n-t)$ submatrix $\bH'$ consisting of the columns
  corresponding to $\cI'$.  Then $\bH'$ is a parity-check matrix for
  $\cC'$, and \[\cK(\bH') = \{\bldx|_{\cI'} \mid \bldx\in\cK(\bH),\
  \bldx|_{\cI\setminus\cI'}=\bldzero\} \:.\] Therefore,
  $\weight_{\min}(\bH')\ge \weight_{\min}(\bH)$, and this proves
  $\rho(\cC')\le\rho(\cC)$.

  For the second inequality, let $\bH'$ be a $\rho'\times(n-t)$
  parity-check matrix for $\cC'$.  Now we consider a $(\rho'+t)\times
  n$ matrix $\bH$ with the following properties: The upper
  $\rho'\times n$ submatrix of $\bH$ consists of the columns of $\bH'$
  at positions $\cI'$ and of zero-columns at positions
  $\cI\setminus\cI'$, and the lower $t\times n$ submatrix consists of
  rows of weight~$1$ that have $1$s at the positions
  $\cI\setminus\cI'$.  Then $\cC=\ker\bH$ and
  \[ \cK(\bH) = \{ \bldx\in\rr^\cI \mid \bldx|_{\cI'}\in\cK(\bH'),\
  \bldx|_{\cI\setminus\cI'}=\bldzero\}\:. \] Consequently,
  $\weight_{\min}(\bH) = \weight_{\min}(\bH')$, and this proves
  $\rho(\cC)\le \rho(\cC')+t$.
\end{IEEEproof}

\begin{lemma}\label{lemma:distance_two}
  Let $\cC$ be a code of minimum distance $d\le 2$.  Then
  $d=\weight_{\min}(\bH)$ for any parity-check matrix $\bH$ of $\cC$,
  i.e., $\cC$ is in class $3$ (for BEC, AWGNC, BSC, and max-fractional
  pseudoweight).
\end{lemma}

\begin{IEEEproof}
  By Lemma~\ref{lemma:relations} it suffices to prove this lemma for
  the max-fractional weight $\weight = \weight_{\maxfrac}$.  Since
  $\weight(\bldx)\ge 1$ holds for all nonzero pseudocodewords, we
  always have ${\weight_{\min}(\bH)\ge 1}$, which proves the result in
  the case $d=1$.

  Let $d=2$ and $\bH$ be a parity-check matrix for $\cC$.  Let
  $\bldx\in\cK(\bH)$ and let $x_{\ell}$ be the largest coordinate.
  Since $d=2$ there is no zero column in $\bH$ and thus there exists a
  row $j$ with $\ell\in\cI_j$.  Then $x_{\ell}\le
  \sum_{i\in\cI\setminus\{\ell\}}x_i$, hence
  $2x_{\ell}\le\sum_{i\in\cI}x_i$, and thus $\weight(\bldx)\ge 2$.  It
  follows $\weight_{\min}(\bH)\ge 2$ and the lemma is proved.
\end{IEEEproof}


\section{Constructions of codes from other codes}%
\label{sec:inheritance}

The following results consider the pseudoredundancy of codes obtained
from other codes by the direct sum or the $(\bldu\,\bldu)$
construction.  They are analogs of Theorems~7 and 8
in~\cite{Schwartz_Vardy}, and Theorems~4.1 and 4.2
in~\cite{Kelley_Sridhara}, for the case of the max-fractional weight
and the AWGNC pseudoweight. Our proofs in each case follow the
exposition of these earlier proofs.

\begin{theorem}\label{thm:concatenation}
  Let $\cC_1$ and $\cC_2$ be $[n_1,k_1,d_1]$ and $[n_2,k_2,d_2]$
  binary linear codes, respectively. Then the direct sum $\cC_3 = \{
  (\bldu \; \bldv) \mid \bldu \in \cC_1, \bldv \in \cC_2 \}$ is an
  $[n_1 + n_2,k_1 + k_2,\min \{ d_1,d_2 \} ]$ code with
  \begin{gather*}
    \rho_{\maxfrac}(\cC_3) \le \rho_{\maxfrac}(\cC_1) +
    \rho_{\maxfrac}(\cC_2) \:, \\
    \rho_{\AWGNC}(\cC_3) \le \rho_{\AWGNC}(\cC_1) +
    \rho_{\AWGNC}(\cC_2) \:.
  \end{gather*}
\end{theorem}

\begin{IEEEproof}
  Without loss of generality, we may assume that both $\rho(C_1)$ and
  $\rho(C_2)$ are finite, for otherwise the statement to be proved is
  trivial.  For $i=1,2$, let $\bH_i$ be a parity-check matrix for
  $\cC_i$ having $\rho(\cC_i)$ rows and such that $\weight(\bldx) \ge
  d_i$ for all $\bldx \in \cK(\bH_i) \setminus \{ \bldzero \}$. Then
  \[ \bH_3 = \left[ \begin{array}{cc}
      \bH_1 & \bldzero \\
      \bldzero & \bH_2 
    \end{array} \right] \]
  is a parity-check matrix for $\cC_3$ with $\rho(\cC_1) +
  \rho(\cC_2)$ rows. Let $\bldp = (\bldq \; \bldr) \in \cK( \bH_3 )
  \setminus \{ \bldzero \}$, where the vectors $\bldq$ and $\bldr$ in
  the concatenation have lengths $n_1$ and $n_2$ respectively. Then,
  we may assume $\bldq \in \cK( \bH_1 )\setminus \{ \bldzero \}$ 
  and $\bldr \in \cK( \bH_2 )\setminus \{ \bldzero \}$, and therefore 
  $\weight(\bldq) \ge d_1$ and $\weight(\bldr) \ge d_2$. 
  (Note that in the case where either $\bldq$ or $\bldr$ is equal to
  $\bldzero$, the result is trivial since for any $\bldq \ne
  \bldzero$, $\weight(\bldq \; \bldzero) = \weight(\bldq)$ 
  for both the max-fractional weight and the AWGNC pseudoweight.)

  We consider the two cases of max-fractional weight and
  AWGNC pseudoweight separately. 

  \emph{Max-fractional weight:} Assume without loss of generality that
  $\max \{ q_i \} \ge \max \{ r_i \}$. Then
  \begin{multline*}
    \weight_{\maxfrac}(\bldp) = \frac{\sum p_i}{\max \{ p_i \} } 
    = \frac{\sum q_i + \sum r_i}{\max \{ q_i \} } \\
    > \frac{\sum q_i}{\max \{ q_i \} } \ge d_1 \ge \min \{ d_1, d_2 \}
  \end{multline*}
  which proves the result. 
  
  \emph{AWGNC pseudoweight:} Assume without loss of generality that
  $\weight_{\AWGNC}(\bldq) \ge \weight_{\AWGNC}(\bldr)$; this
  condition may be written as
  \begin{equation}
    \bigg( \sum_{i=1}^{n_1} q_i \bigg)^{\!2} \bigg( \sum_{i=1}^{n_2} 
    r_i^2 \bigg) \ge \bigg( \sum_{i=1}^{n_1} q_i^2 \bigg) 
    \bigg( \sum_{i=1}^{n_2} r_i \bigg)^{\!2}  \:.
    \label{eq:ineq_1_thm_AWGNC}
  \end{equation}
  To establish the result, we need only to prove that
  $\weight_{\AWGNC}(\bldp) \ge \weight_{\AWGNC}(\bldr)$. Now, since
  the entries of $\bldq$ and $\bldr$ are nonnegative, we have
  \begin{equation}
    2 \bigg( \sum_{i=1}^{n_1} q_i \bigg) \bigg( \sum_{i=1}^{n_2} r_i 
    \bigg) \bigg( \sum_{i=1}^{n_2} r_i^2 \bigg) \ge 0 \; .
    \label{eq:ineq_2_thm_AWGNC}
  \end{equation}
  Adding $\big( \sum_{i=1}^{n_2} r_i \big)^2 \big( \sum_{i=1}^{n_2}
  r_i^2 \big)$ to both sides of \eqref{eq:ineq_2_thm_AWGNC} and adding
  the resulting inequality to inequality \eqref{eq:ineq_1_thm_AWGNC}
  yields
  \[
  \bigg( \sum_{i=1}^{n_1} q_i + \sum_{i=1}^{n_2} r_i \bigg)^{\!\!2} 
  \bigg( \sum_{i=1}^{n_2} r_i^2 \bigg) 
  \ge \bigg( \sum_{i=1}^{n_2} r_i \bigg)^{\!\!2} \bigg( \sum_{i=1}^{n_1} 
  q_i^2 + \sum_{i=1}^{n_2} r_i^2 \bigg)
  \]
  which may be rearranged as $\weight_{\AWGNC}(\bldp) \ge
  \weight_{\AWGNC}(\bldr)$, as desired.
\end{IEEEproof}

\begin{theorem}\label{thm:self-concatenation}
  Let $\cC_1$ be an $[n,k,d]$ binary linear code. Then $\cC_2 = \{
  (\bldu \; \bldu) \mid \bldu \in \cC_1 \}$ is a $[2n, k, 2d]$ code with
  \begin{gather*}
    \rho_{\maxfrac}(\cC_2) \le \rho_{\maxfrac}(\cC_1) + n \:, \\
    \rho_{\AWGNC}(\cC_2) \le \rho_{\AWGNC}(\cC_1) + n \:.
  \end{gather*}
\end{theorem}

\begin{IEEEproof}
  As before, without loss of generality, we may assume that
  $\rho(C_1)$ is finite.  Let $\bH_1$ be a parity-check matrix for
  $\cC_1$ with $\rho(\cC_1)$ rows and such that $\weight(\bldx) \ge
  d_1$ for all $\bldx \in \cK(\bH_1) \setminus \{ \bldzero \}$. Then
  \[ \bH_2 = \left[ \begin{array}{cc}
      \bH_1 & \bldzero \\
      \bI_n & \bI_n 
    \end{array} \right] \]
  is a parity-check matrix for $\cC_2$ with $\rho(\cC_1) + n$ rows
  (here $\bI_n$ denotes the $n \times n$ identity matrix). Let $\bldp
  = (\bldq \; \bldr) \in \cK(\bH_2) \setminus \{ \bldzero \}$, where
  the vectors $\bldq \in \cK(\bH_1)$ and $\bldr$ in the concatenation
  both have length $n$. Then, for $i=1,2,\ldots,n$, from the
  fundamental cone inequalities for row $n+i$ we get $q_i \le r_i \le
  q_i$, so we have $\bldp = (\bldq \; \bldq)$. Now, since $\bldq \in
  \cK(\bH_1) \setminus \{ \bldzero \}$, we have $\weight(\bldq) \ge
  d_1$. Since $\weight((\bldq \; \bldq)) = 2\, \weight(\bldq)$ for
  both the max-fractional weight and the AWGNC pseudoweight, we have
  $\weight(\bldp) \ge 2d$, and the result follows.
\end{IEEEproof}

\begin{remark}
  Theorem~9 in~\cite{Schwartz_Vardy} and Theorem~4.3
  in~\cite{Kelley_Sridhara} state that if $\cC$ is an $[n,k,3]$ binary
  linear code then the extended $[n\!+\!1,k,4]$ code $\cC'$ satisfies
  $\rho(C')\le 2\rho(\cC)$, for the BEC pseudoweight and the BSC
  pseudoweight, respectively.  Regarding the corresponding results for
  the case of the max-fractional weight and the AWGNC pseudoweight, we
  mention here only that the analogous result in fact does not hold
  for the case of the max-fractional weight. As a counterexample,
  consider the $[7,4,3]$ Hamming code $\cC_1$ which satisfies
  $\rho(\cC_1) \le 2^3-1 = 7$ (cf.\ Proposition
  \ref{prop:Hamming}). On the other hand, the $[8,4,4]$ extended
  Hamming code $\cC_2$ satisfies $\rho_{\maxfrac}(\cC_2) = \infty$
  (cf.\ Section \ref{sec:short_codes_results}).
\end{remark}


\section{Parity-check matrices with rows of weight~$2$}%
\label{sec:row_weight_2}

In this section we consider the pseudoredundancy of codes with a
parity-check matrix consisting of rows of weight~$2$ and at most one
additional row.  The results are then applied to upper-bound the
pseudoredundancy for codes of dimension~$2$.  The basic case is dealt
with in the following lemma.

\begin{lemma}\label{lemma:row_weight_2}
  Let $\bH$ be a parity-check matrix of $\cC$ such that every row in
  $\bH$ has weight~$2$.  Then:
  \begin{enumerate}
  \item[(a)] There is an equivalence relation on the set $\cI$ of column
    indices of $\bH$ such that for a vector $\bldx\in\rr^n$ with
    non-negative coordinates we have $\bldx\in\cK(\bH)$ if and only if
    $\bldx$ has equal coordinates within each equivalence class.
  \item[(b)] The minimum distance of $\cC$ is equal to its minimum
    BEC, AWGNC, BSC, and max-fractional pseudoweights with respect to
    $\bH$, i.e., $d(\cC) = \weight_{\min}(\bH)$.
  \end{enumerate}
\end{lemma}

\begin{IEEEproof} 
  For (a), define the required relation $R$ as follows: For
  $i,i'\in\cI$ let $(i,i')\in R$ if and only if $i=i'$ or there exists
  an integer $\ell \ge 1$, column indices $i=i_0, i_1, \dots,
  i_{\ell-1}, i_{\ell}=i'\in\cI$ and row indices $j_1,\dots,j_l\in\cJ$
  such that
    \[ \{ i_0, i_1 \} = \cI_{j_1} \, , \,
    \{ i_1, i_2 \} = \cI_{j_2} \, , \, \dots \, , \,
    \{ i_{\ell-1}, i_{\ell} \} = \cI_{j_{\ell}} \; . \]
  This is an equivalence relation, and it defines equivalence
  classes over $\cI$.  It is easy to check that
  inequalities~(\ref{eq:fundcone-inequality-1}) imply that
  $\bldx\in\cK(\bH)$ if and only if $x_i=x_{i'}$ for any $(i,i')\in
  R$.

  In order to prove (b), we note that the minimum (BEC, AWGNC, BSC or
  max-fractional) pseudoweight is always bounded above by the minimum
  distance of $\cC$, so we only have to show that the minimum
  pseudoweight is bounded below by the minimum distance.
    
  Let $\cS = \{S_1, S_2, \dots, S_t\}$ be the set of equivalence
  classes of $R$, and let $d_S = |S|$ for $S\in\cS$.  It is easy to
  see that the minimum distance of $\cC$ is $d = \min_{S\in\cS}d_S$
  (since the minimum weight nonzero codeword of $\cC$ has non-zeros in
  the coordinates corresponding to a set $S\in\cS$ of minimal size and
  zeros everywhere else).

  Now let $\bldx\in\cK(\bH)$.  Since the coordinates $x_i$, $i\in\cI$,
  depend only on the equivalence classes, we may use the notation
  $x_S$, $S\in\cS$.  Let $x_T$, $T\in\cS$, be the largest
  coordinate. Then: 
  \[ \weight_{\maxfrac}(\bldx) = \frac{\sum_{i \in \cI} x_i}{x_T} \ge
  \frac{\sum_{i\in T}x_i}{x_T} = |T|=d_T \ge d \:. \] Therefore,
  $\weight_{\min}^{\maxfrac} (\bH) \ge d$, and by using
  Lemma~\ref{lemma:relations}, we get
  $\weight_{\min}^{\BEC}(\bH)\ge d$, $\weight_{\min}^{\AWGNC}(\bH)\ge
  d$, and $\weight_{\min}^{\BSC}(\bH)\ge d$.
\end{IEEEproof}

The following proposition is a stronger version of
Lemma~\ref{lemma:row_weight_2}.

\begin{proposition}\label{prop:row_weight_2}
  Let $\bH$ be an $m \times n$ parity-check matrix of $\cC$, and
  assume that the $m\!-\!1$ first rows in $\bH$ have weight~$2$.
  Denote by $\widehat{\bH}$ the $(m\!-\!1) \times n$ matrix consisting
  of these rows, consider the equivalence relation of
  Lemma~\ref{lemma:row_weight_2} (a) with respect to $\widehat{\bH}$,
  and assume that $\cI_m$ intersects each equivalence class in at most
  one element.  Then, the minimum distance of $\cC$ is equal to its
  minimum BEC, AWGNC, BSC, and max-fractional pseudoweights with
  respect to $\bH$, i.e., $d(\cC) = \weight_{\min}(\bH)$.
\end{proposition}

\begin{IEEEproof} 
  Let $\cS$ be the set of classes of the aforementioned equivalence
  relation on $\cI$, and let $d_S=|S|$ for $S\in\cS$.  Let
  \[ \cS' = \{ S \in \cS \mid\: |S \cap \cI_m| = 1 \} \:.  \] Also
  let $\cS'' = \cS \setminus \cS'$, so that $S \cap \cI_m =
  \varnothing$ for all $S \in \cS''$.

  Let $\bldx\in\cK(\bH) \setminus \{ \bldzero \}$. As before, since
  the coordinates $x_i$, $i\in\cI$, depend only on the equivalence
  classes, we may use the notation $x_S$, $S\in\cS$.  The fundamental
  cone constraints \eqref{eq:fundcone-inequality-1} and
  \eqref{eq:fundcone-inequality-2} may then be written as $x_S \ge 0$
  for all $S \in \cS$ and
  \begin{equation}
    \forall R \in \cS' :\  x_R \le \sum_{S \in \cS'\setminus\{R\}} x_S \; , 
    \label{eq:fundcone_ineq_S'}
  \end{equation}
  respectively, and the max-fractional weight of $\bldx \in \cK(\bH)
  \setminus \{ \bldzero \}$ is given by
  \begin{equation}
    \weight_{\maxfrac}(\bldx) 
    = \frac{\sum_{S \in \cS} d_S x_S}{\max_{S \in \cS} x_S} \; .
    \label{eq:new_maxfrac}
\end{equation}

Suppose $\bldx\in\cK(\bH) \setminus \{ \bldzero \}$ has minimal
max-fractional weight.  Let $x_T$ be its largest coordinate. First
note that if there exists $R \in \cS''\setminus\{T\}$ with $x_R > 0$,
setting $x_R$ to zero results in a new pseudocodeword with lower
max-fractional weight, which contradicts the assumption that $\bldx$
achieves the minimum. Therefore $x_R = 0$ for all $R \in
\cS''\setminus\{T\}$. We next consider two cases.

\emph{Case 1:} $T \in \cS''$. If there exists $R \in \cS'$ with $x_R >
0$, setting all such $x_R$ to zero results in a new pseudocodeword
with lower max-fractional weight, which contradicts the minimality of
the max-fractional weight of $\bldx$. Therefore $x_T$ is the only
positive coordinate of $\bldx$, and by \eqref{eq:new_maxfrac} the
max-fractional weight of $\bldx$ is $d_T$.

\emph{Case 2:} $T \in \cS'$. In this case $x_R = 0$ for all $R \in
\cS''$. From inequality~(\ref{eq:fundcone_ineq_S'}) for $R=T$ we
obtain
\[ x_T \le \sum_{S\in\cS'\setminus\{T\}} x_S \:. \]
With $d_0 \define \min_{S\in\cS'\setminus\{T\}}d_S$ it follows that
\[ d_0 x_T \le \sum_{S\in\cS'\setminus\{T\}} d_0 x_S \le
\sum_{S\in\cS'\setminus\{T\}} d_Sx_S \:. \] Consequently, \[
(d_T+d_0)x_T \le \sum_{S\in\cS} d_Sx_S \:, \] and thus
$\weight_{\maxfrac}(\bldx) \ge d_T+d_0$.  We conclude that the minimum
max-fractional weight is given by
\[ \weight_{\min}^{\maxfrac} (\bH) = \min \left\{ \min_{S, T \in \cS',
    S \neq T} \{d_S + d_T\} \; , \; \min_{S \in \cS''} \{d_S \}
\right\} \:. \] But this is easily seen to be equal to the minimum
distance $d$ of the code.

Finally, by using Lemma~\ref{lemma:relations}, we obtain that
$\weight_{\min}^{\BEC}(\bH)=d$, $\weight_{\min}^{\AWGNC}(\bH)=d$ and
$\weight_{\min}^{\BSC}(\bH)=d$.
\end{IEEEproof}

\begin{remark}
  The requirement that all $i \in \cI_m$ belong to different
  equivalence classes of $\widehat{\bH}$ in
  Proposition~\ref{prop:row_weight_2} is necessary. Indeed, consider
  the matrix
  \[ \bH = \mat{
    \ze & \ze & \zo & \zo \\
    \zo & \ze & \ze & \zo \\
    \ze & \zo & \ze & \zo \\
    \ze & \ze & \ze & \ze \\
  } \:. \]
  One can see that there are two equivalence classes for
  $\widehat{\bH}$: $S_1 = \{ 1, 2, 3 \}$, $S_2 = \{ 4 \}$. The minimum
  distance of the corresponding code $\cC$ is $4$ (since $(1, 1, 1, 1)$
  is the only nonzero codeword).  However, $\bldx = (1, 1, 1, 3) \in
  \cK(\bH)$ is a pseudocodeword of max-fractional weight $2$.
\end{remark}
  
\begin{corollary}\label{corollary:dimension_two}
  Let $\cC$ be a code of length~$n$ and dimension~$2$.  Then
  $\rho(\cC)=n-2$, i.e., $\cC$ is of class at least~$2$ (for BEC,
  AWGNC, BSC, and max-fractional  pseudoweight).
\end{corollary}\pagebreak

\begin{IEEEproof}
  We consider two cases. 

  {\em Case 1: $\cC$ has no zero coordinates.}
  
  Let $\bldc_1$ and $\bldc_2$ be two linearly independent codewords
  of $\cC$.  Define the following subsets of $\cI$:
  \begin{eqnarray*}
    S_1 & \define &  \{ i \in \cI \mid 
    i \in \mbox{supp}(\bldc_1) 
    \mbox{ and } i \notin \mbox{supp}(\bldc_2) \} \; \\
    S_2 & \define & 
    \{ i \in \cI \mid 
    i \notin \mbox{supp}(\bldc_1) 
    \mbox{ and } i \in \mbox{supp}(\bldc_2) \} \; \\
    S_3 & \define & \{ i \in \cI \mid 
    i \in \mbox{supp}(\bldc_1) 
    \mbox{ and } i \in \mbox{supp}(\bldc_2) \} .
  \end{eqnarray*}
  The sets $S_1$, $S_2$, and $S_3$ are pairwise disjoint.  Since $\cC$
  has no zero coordinates, $\cI = S_1 \cup S_2 \cup S_3$.  The
  ordering of the elements in $\cI$ implies an ordering on the
  elements in each of $S_1$, $S_2$, and $S_3$.  Assume that $S_1 = \{
  i_1, i_2, \cdots, i_{|S_1|}\}$ and $i_1 < i_2 < \cdots < i_{|S_1|}$.
  If $S_1 \neq \varnothing$, let $m_1 = i_1$ be the minimal element in
  $S_1$, and define an $(|S_1| - 1) \times n$ matrix $\bH_1 =
  (H^1_{j,\ell})$ as follows:
  \[ H^1_{j,\ell} = \left\{ \begin{array}{cl}
      1 & \mbox{ if } i_j = \ell \mbox{ or } i_{j+1} = \ell \; , \\
      & \qquad \qquad j = 1, 2, \cdots, |S_1|-1 \; , \\
      0 & \mbox{ otherwise } \; .  
    \end{array} \right.  \]
  Similarly, define $(|S_2| - 1) \times n$ and $(|S_3| - 1) \times n$
  matrices $\bH_2$ and $\bH_3$, with respect to $S_2$ and $S_3$. 
  (Some of the $S_i$s might be equal to $\varnothing$, in which case
  the corresponding $\bH_i$ is not defined.) Let $m_2$
  and $m_3$ be minimal elements of $S_2$ and $S_3$, respectively (if $S_2 
  \neq \varnothing$ and $S_3 \neq \varnothing$).
  
  {\em Subcase 1-a: One of $S_1$, $S_2$, $S_3$ is empty.}  Without
  loss of generality we may assume that $S_3 = \varnothing$, i.e.,
  that $c_1$ and $c_2$ have disjoint support; indeed, if for example
  $S_1 = \varnothing$, then $\mathrm{supp}(c_1) \subseteq
  \mathrm{supp}(c_2)$ and we can replace $c_2$ by $c_1 + c_2$.  Define
  an $(n-2) \times n$ matrix $\bH$ by $\bH^T \define [ \, \bH_1^T \; |
  \; \bH_2^T \,]$.  It is easy to see that all rows of $\bH$ are
  linearly independent, and so its rank is $n-2$. It is also
  straightforward that for all $\bldc \in \cC$ we have $\bldc \in
  \ker(\bH)$. Therefore, $\bH$ is a parity-check matrix of~$\cC$. The
  matrix $\bH$ has a form as in Lemma~\ref{lemma:row_weight_2}, and
  thus $\rho(\cC) = n-2$.
 
  {\em Subcase 1-b: Neither of $S_1$, $S_2$, $S_3$ is empty.}  
  Define a $1 \times n$ matrix $\bH_4 = (H^4_{j,\ell})$, where
  \[ H^4_{1,\ell} = \left\{ \begin{array}{cl}
      1 & \mbox{ if } S_j \neq \varnothing \mbox{ and } m_j = \ell \\
      & \qquad \qquad \mbox{ for } j = 1, 2, 3 \; , \\
      0 & \mbox{ otherwise } \; .  
    \end{array} \right.  \]
  
  Additionally, define an $(n-2) \times n$ matrix $\bH$ by $\bH^T
  \define [ \, \bH_1^T \; | \; \bH_2^T \; | \; \bH_3^T \; | \; \bH_4^T
  \,]$.  Similarly to the previous case, all rows of $\bH$ are
  linearly independent, its rank is $n-2$. For all $\bldc \in \cC$ we
  have $\bldc \in \ker(\bH)$.  Therefore, $\bH$ is a parity-check
  matrix of~$\cC$.
  
  The matrix $\bH$ has a form as in
  Proposition~\ref{prop:row_weight_2} (where $S_1$, $S_2$, and $S_3$
  are corresponding equivalence classes over $\cI$), and therefore
  $\rho(\cC) = n-2$.
  
  {\em Case 2: $\cC$ has $t>0$ zero coordinates.}
  
  Consider a code $\cC'$ of length $n-t$ obtained by puncturing~$\cC$
  in these $t$ zero coordinates. From Case 1 (with respect to $\cC'$),
  $\rho(C') = n-t-2$. By applying the rightmost inequality in
  Lemma~\ref{lemma:puncturing}, we have $\rho(C) \le n-2$. Since
  $k=2$, we conclude that $\rho(C) = n-2$.
\end{IEEEproof}


\section{Codes Based on Designs}\label{sec:designs}

Among the codes with finite pseudoredundancy an interesting class of
codes is based on designs.  In this section we consider {\em partial
  designs}, which include the common BIBDs (also called $2$-designs).
We present a principal lower bound on the minimum pseudoweight for
codes, when the parity-check matrix is the block-point incidence
matrix of a partial design.  We apply this bound to the Hamming codes
and the simplex codes and deduce that their pseudoredundancy is
finite.

\begin{definition}\label{prop:design_incidence_matrix}
  A \emph{partial $(w_c,\lambda)$ design} is a block design consisting
  of an $n$-element set $\cV$ (whose elements are called
  \emph{points}) and a collection of $m$ subsets of $\cV$ (called
  \emph{blocks}) such that every point is contained in exactly $w_c$
  blocks and every $2$-element subset of $\cV$ is contained in at most
  $\lambda$ blocks. The \emph{incidence matrix} of a design is an $m
  \times n$ matrix $\bH$ whose rows correspond to the blocks and whose
  columns correspond to the points, and that satisfies $H_{j,i} = 1$
  if block $j$ contains point~$i$, and $H_{j,i} = 0$ otherwise.

  If each block contains the same number $w_r$ of points and every
  $2$-element subset of $\cV$ is contained in exactly $\lambda$
  blocks, the design is said to be an $(n,w_r,\lambda)$ \emph{balanced
    incomplete block design} (BIBD), or \emph{$2$-design}.
\end{definition}

In the following we avoid the trivial cases $n\le 1$ and $\lambda=0$.
For a BIBD we have $n\, w_c = m\, w_r$ and also
\[ w_c\,(w_r-1) = \lambda\,(n-1) \]
(see, e.g., \cite[p.~60]{MacWilliams_Sloane}), so $(n,w_r,\lambda)$
determines the other parameters $w_c$ and $m$ by
\begin{equation}\label{eq:design_constraint}
  w_c = \frac{n-1}{w_r-1}\,\lambda
  \quad\text{and}\quad
  m = \frac{n\,(n-1)}{w_r\,(w_r-1)}\,\lambda \:.
\end{equation}
Note that \cite{Vasic_Milenkovic} and \cite{Kashyap_Vardy} consider
parity-check matrices based on BIBDs; these matrices are the transpose
of the incidence matrices defined here.

We have the following general result for codes based on partial
$(w_c,\lambda)$ designs.

\begin{theorem}\label{thm:pseudoweight_bound}
  Let $\cC$ be a code with parity-check matrix $\bH$, such that a
  subset of the rows of $\bH$ forms the incidence matrix for a partial
  $(w_c,\lambda)$ design. Then the minimum max-fractional weight of
  $\cC$ with respect to $\bH$ is lower bounded by
  \begin{equation}
    \weight_{\min}^{\maxfrac} \ge 1 + \frac{w_c}{\lambda} \; .
    \label{eq:new_lower_bound}
  \end{equation}
  For the case of an $(n,w_r,\lambda)$ BIBD, the lower bound in
  (\ref{eq:new_lower_bound}) may also be written as 
  \[ \weight_{\min}^{\maxfrac} \ge 1 + \frac{n-1}{w_r-1} \:; \]
  the alternative form follows directly from
  (\ref{eq:design_constraint}).
\end{theorem}

\begin{IEEEproof}
  Consider the subset of the rows of $\bH$ which forms the incidence
  matrix for a partial $(w_c,\lambda)$ design.  Let $\bldx$ be a
  nonzero pseudocodeword and let $x_\ell$ be a maximal coordinate of
  $\bldx$ ($\ell \in \cI$).  For all $j \in \cJ_\ell$, sum
  inequalities (\ref{eq:fundcone-inequality-1}). We have
  \begin{equation*}
    w_c x_{\ell} \le \lambda\! \sum_{i \in \cI \setminus \{ \ell \} } x_i \; ,
  \end{equation*}
  and thus
  \begin{equation}\label{eq:basic_step_for_bound_bsc} 
    \left( 1 + \frac{w_c}{\lambda} \right) x_{\ell} \le 
    \sum_{i \in \cI} x_i \; . 
  \end{equation}
  The result now easily follows from the definition of
  $\weight_{\min}^{\maxfrac}$.
\end{IEEEproof}

\begin{theorem}\label{thm:pseudoweight_bound_2}
  Let $\cC$ be a code with parity-check matrix $\bH$, such that a
  subset of the rows of $\bH$ forms the incidence matrix for a partial
  $(w_c,\lambda)$ design. Then,
  \begin{align*}
    \weight_{\min}^{\AWGNC} & \ge 1 + \frac{w_c}{\lambda} \; , \\
    \weight_{\min}^{\BSC} & \ge 1 + \frac{w_c}{\lambda} \; . 
  \end{align*}
\end{theorem}

\begin{IEEEproof}
  Apply Lemma~\ref{lemma:relations} and
  Theorem~\ref{thm:pseudoweight_bound}.
\end{IEEEproof}

Results similar to Theorem~\ref{thm:pseudoweight_bound} and
Theorem~\ref{thm:pseudoweight_bound_2} were also presented and proven
by Xia and Fu \cite{Xia_Fu} in the AWGNC case.

\begin{remark}\label{rem:spectrum_gap}
  Under the conditions of Theorem~\ref{thm:pseudoweight_bound_2}, if
  $\bldx\in\cK(\bH)$ is a nonzero pseudocodeword such that
  $\weight_{\AWGNC}(\bldx) = 1+\frac{w_c}{\lambda}$ holds then it
  follows that $\bldx$ is a scalar multiple of a binary vector.  This
  can be easily seen by considering the proof of the inequality
  $\weight_{\AWGNC}(\bldx)\ge \weight_{\maxfrac}(\bldx)$ (see
  \cite[Lemma 44]{KV-long-paper}) and examining when equality
  $\weight_{\AWGNC}(\bldx) = \weight_{\maxfrac}(\bldx)$ holds.

  Furthermore it can be shown that in this case $\bldx$ is actually a
  scalar multiple of a codeword (see \cite[Theorem 3]{Xia_Fu}).  It
  follows that the AWGNC pseudocodeword spectrum gap is positive,
  provided that $d(\cC) = 1+\frac{w_c}{\lambda}$ holds.
\end{remark}

Another tool for proving lower bounds on the minimum AWGNC
pseudoweight is provided by the following eigenvalue-based lower bound
by Vontobel and Koetter \cite{KV-lower-bounds}.

\begin{proposition}[cf.\ \cite{KV-lower-bounds}]\label{prop:KV_bound}
  The minimum AWGNC pseudoweight for a $(w_c,w_r)$-regular
  parity-check matrix $\bH$ whose corresponding Tanner graph is
  connected is bounded below by
  \begin{equation}\label{eq:KV_bound}
    \weight_{\min}^{\AWGNC} \ge n \cdot \frac{ 2w_c - \mu_2 } 
    {\mu_1 - \mu_2 } \; ,
  \end{equation}
  where $\mu_1$ and $\mu_2$ denote the largest and second largest
  eigenvalue (respectively) of the matrix $\bL \define \bH^T \bH$;
  here, $\bL$ and the matrix multiplication are to be considered over
  the reals.
\end{proposition}

In the case where $\bH$ is equal to the incidence matrix for an
$(n,w_r,\lambda)$ BIBD, the bound of Proposition \ref{prop:KV_bound}
becomes
\begin{equation}\label{eq:KV_bound_2}
  \weight_{\min}^{\AWGNC} \ge 1 + \frac{ w_c }{ \lambda } \; ,
\end{equation}
so that in this case the bound of Proposition~\ref{prop:KV_bound}
coincides with that of Theorem \ref{thm:pseudoweight_bound_2} (for the
case of the AWGNC only).

To see why (\ref{eq:KV_bound}) becomes (\ref{eq:KV_bound_2}), denote
the column $i\in\cI$ of $\bH$ by $\bldh_i$ and denote the matrix $\bL
= \left( L_{i,\ell} \right)_{i,\ell \in \cI} = \bH^T \bH$.  From the
properties of a BIBD we get
\[ L_{i,\ell} = \bldh_i^T \bldh_{\ell} = \begin{cases}
  w_c & \text{if } i = \ell \:, \\
  \lambda & \text{if } i \ne \ell \:.
\end{cases} \] Now, $\bL$ has largest eigenvalue $\mu_1 = w_r w_c$ and
only one other eigenvalue $\mu_2 = w_c-\lambda$, whose multiplicity is
$n-1$, since one can write $\bL = \lambda\bldone + (w_c-\lambda)
\bldI$, where $\bldone$ and $\bldI$ denote the all-ones and the
identity matrices, respectively.  Now we have $2w_c-\mu_2 =
w_c+\lambda$ and $\mu_1-\mu_2 = w_rw_c - w_c + \lambda = n\lambda$, so
that $n \cdot \frac{ 2w_c - \mu_2 } {\mu_1 - \mu_2 } = 1 +
\frac{w_c}{\lambda}$.\smallskip

\begin{remark}
  Prominent examples for codes based on designs are codes based on
  Euclidean or projective geometries, in particular the $[{4^s-1}\,,
  {4^s-3^s}\,, {2^s+1}]$ code based on the Euclidean plane ${\rm
    EG}(2, 2^s)$ as well as the $[{4^s+2^s+1}\,, {4^s-3^s+2^s}\,,
  {2^s+2}]$ code based on the projective plane ${\rm PG}(2, 2^s)$
  (see~\cite{Kou_Lin_Fossorier, Smarandache_Vontobel}).
  Theorem~\ref{thm:pseudoweight_bound_2} and
  Remark~\ref{rem:spectrum_gap} apply to these codes, as their
  standard parity-check matrices form the incidence matrix for a
  partial design with parameters $(w_c,\lambda)=(2^s,1)$ and
  $(w_c,\lambda)=({2^s+1}\,,1)$, respectively\footnote{In the latter
    case the partial design is even a BIBD with parameters
    $({4^s\!+\!2^s\!+\!1}\,, {2^s\!+\!1}\,, 1)$.}; in particular these
  codes have finite pseudoredundancy.
\end{remark}

We next apply the bounds of Theorems~\ref{thm:pseudoweight_bound}
and~\ref{thm:pseudoweight_bound_2} to some other examples of codes
derived from designs.

\begin{proposition}\label{prop:Hamming}
  For $m \ge 2$, the $[{2^m-1}\,, {2^m-1-m}\,, 3]$ Hamming code has
  BEC, AWGNC, BSC, and max-fractional pseudocodeword redundancies \[
  \rho(\cC) \le 2^m-1 \:. \]
\end{proposition}

\begin{IEEEproof}
  For $m \ge 2$, consider the binary parity-check matrix $\bH$ whose
  rows are exactly the nonzero codewords of the dual code $\cC\dual$,
  in this case the $[{2^m-1}\,,m\,,{2^{m-1}}\,]$ simplex code. This
  $\bH$ is the incidence matrix for a BIBD with parameters
  $(n,w_r,\lambda)$ = $({2^m-1}\,, {2^{m-1}}\,, {2^{m-2}})$.
  Theorem~\ref{thm:pseudoweight_bound} gives $\weight_{\maxfrac}
  (\bldx) \ge 3$, leading to $\rho_{\maxfrac} (\cC) \le 2^m-1$.

  The result for BEC, AWGNC, and BSC follows by applying
  Theorem~\ref{thm:pseudoredundancies}.
\end{IEEEproof}

In the next example, we consider simplex codes. Straightforward
application of the previous reasoning does not lead to the desired
result.  However, more careful selection of the matrix $\bH$, as
described below, leads to a new bound on the pseudoredundancy.

\begin{proposition}\label{prop:simplex}
  For $m \ge 2$, the $[{2^m-1}\,, m\,, {2^{m-1}}]$ simplex code has
  BEC, AWGNC, BSC, and max-fractional pseudocodeword redundancies
  \begin{eqnarray*}
    \rho(\cC) \le \frac{(2^m-1)\,(2^{m-1}-1)}{3} \; . 
  \end{eqnarray*}
\end{proposition}

\begin{IEEEproof}
  For $m \ge 2$, consider the binary parity-check matrix $\bH$ whose
  rows are exactly the codewords of the dual code $\cC\dual$ (in this
  case the $[{2^m-1}\,, {2^m-1-m}\, ,3]$ Hamming code) with Hamming
  weight equal to $3$. This $\bH$ is the incidence matrix for a BIBD
  with parameters $(n,w_r,\lambda)$ = $({2^m-1}\,, 3\,, 1)$.
  Theorem~\ref{thm:pseudoweight_bound} gives
  $\weight_{\min}^{\maxfrac} \ge 2^{m-1}$.

  Note that the number of codewords of weight~$3$ in the $[{2^m-1}\,,
  {2^m-1-m}\,, 3]$ Hamming code equals ${(2^m-1)}{(2^{m-1}-1)}/3$.
  One can show this, e.g., by considering the full sphere-packing of
  the perfect Hamming code and observing that each codeword of weight
  $3$ covers exactly $3$ vectors of weight $2$, of which there are
  $(2^m-1)(2^m-2)/2$ in total.

  Next, we justify the claim that $\bH$ is a parity-check matrix of
  $\cC$.  A theorem of Simonis~\cite{Simonis} states that if there
  exists a linear $[n,k,d]$ code then there also exists a linear
  $[n,k,d]$ code whose codewords are spanned by the codewords of
  weight~$d$.  Since the Hamming code is unique for the parameters
  $[{2^m-1}\,, {2^m-1-m}\,, 3]$, this implies that the Hamming code
  itself is spanned by the codewords of weight $3$, so the rowspace of
  $\bH$ equals $\cC$.

  The result for BEC, AWGNC, and BSC follows again by applying
  Theorem~\ref{thm:pseudoredundancies}.
\end{IEEEproof}

We remark that the bounds of Propositions~\ref{prop:Hamming}
and~\ref{prop:simplex} are sharp at least for the case $m=3$ and the
max-fractional weight, see Section~\ref{sec:short_codes_results}.

The following proposition proves that the AWGNC, BSC, and
max-fractional pseudocodeword redundancies are finite for all codes
$\cC$ with minimum distance at most $3$.

\begin{proposition}\label{prop:D3_codes}
  Let $\cC$ be a $[n,k,d]$ code with $d\le 3$.  Then
  $\rho_{\maxfrac}(\cC)$ is finite. Moreover, we have
  $\rho_{\maxfrac}(\cC) = n-k$ in the case $d \le 2$.
\end{proposition}

\begin{IEEEproof}
  By using Lemma~\ref{lemma:distance_two} we may assume $d=3$.  Denote
  by $\bH$ the parity-check matrix whose rows consist of \emph{all}
  codewords of the dual code of $\cC$. Note that for a code of minimum
  distance $d$, a parity-check matrix $\bH$ consisting of all rows of
  the dual code $\cC\dual$ is an orthogonal array of strength
  $d-1$. In the present case $d=3$, and this implies that in any pair
  of columns of $\bH$, all length-$2$ binary vectors occur with equal
  multiplicities (cf.\ \cite[p. 139]{MacWilliams_Sloane}). Thus the
  matrix $\bH$ is an incidence matrix for a partial block design with
  parameters $(w_c,\lambda) = (2^{r-1},2^{r-2})$, where $r=n-k$.
  Therefore for this matrix $\bH$ the code has minimum (AWGNC, BSC, or
  max-fractional) pseudoweight at least $1 + w_c/\lambda = 3$, and it
  follows that the pseudocodeword redundancy is finite for any code
  with $d=3$.
\end{IEEEproof}

We remark that Proposition~\ref{prop:D3_codes} implies the results
for the Hamming codes (Proposition~\ref{prop:Hamming}). However, we
present the two proofs, since they use different methods.

We have considered in this section several families of codes based on
designs, which have finite pseudocodeword redundancy.  As noted in
Section~\ref{sec:related}, finiteness of pseudoredundancy would also
follow if one can show that the codes are geometrically perfect.
However, this is not the case for the examined codes in general.  For
example, the $[{2^m-1} \,, {2^m-1-m}\,, 3]$ Hamming code is not
geometrically perfect for $m\ge 4$; this follows from the
characterization of geometrically perfect codes, as the $[7,3,4]$
simplex code can be obtained from the Hamming code by repeated
shortening, when $m\ge 4$.


\section{The Pseudocodeword Redundancy for Codes of Small Length}%
\label{sec:short_codes}

In this section we compute the AWGNC, BSC, and max-fractional
pseudocodeword redundancies for all codes of small length.  By
Lemma~\ref{lemma:distance_two} it is sufficient to examine only codes
with minimum distance at least~$3$.  Furthermore, in light of
Lemma~\ref{lemma:puncturing} we will consider only codes without zero
coordinates, i.e., codes that have a dual minimum distance of at
least~$2$.  Finally, we point out to
Corollary~\ref{corollary:dimension_two} for codes of dimension~$2$, by
which we may focus on codes with dimension at least~$3$.

\subsection{The Algorithm}

To compute the pseudocodeword redundancy of a code $\cC$ we have to
examine all possible parity-check matrices for the code $\cC$, up to
equivalence.  Here, we say that two parity-check matrices $\bH$ and
$\bH'$ for the code $\cC$ are \emph{equivalent} if $\bH$ can be
transformed into $\bH'$ by a sequence of row and column permutations.
In this case, $\weight_{\min}(\bH) = \weight_{\min}(\bH')$ holds for
the BEC, AWGNC, BSC, and max-fractional pseudoweights. The enumeration
of codes and parity-check matrices can be described by the following
algorithm.

\begin{algorithm}\hrulefill\smallskip

\textbf{Input:} Parameters $n$~(code length), $k$~(code dimension),
$\rho$~(number of rows of the output parity-check matrices), where
$\rho\ge r\define n-k$.

\textbf{Output:} For all codes of length $n$, dimension $k$, minimum
distance $d\ge 3$, and without zero coordinates, up to code
equivalence: a list of all $\rho\times n$ parity-check matrices, up to
parity-check matrix equivalence.

\begin{enumerate}
\item Collect the set $X$ of all $r\times n$ matrices such that
  \begin{itemize}
  \item they have different nonzero columns, ordered
    lexicographically,
  \item there is no non-empty $\ff_2$-sum of rows which has weight~$0$
    or~$1$ {\sl (this way, the matrices are of full rank and the
      minimum distance of the row space is at least~$2$)}.
  \end{itemize}
\item Determine the orbits in $X$ under the action of the group
  $\GL_r(2)$ of invertible $r\times r$ matrices over $\ff_2$ {\sl
    (this enumerates all codes with the required properties, up to
    equivalence; the codes are represented by parity-check matrices)}.
\item For each orbit $X_{\cC}$, representing a code $\cC$:
  \begin{enumerate}
  \item Determine the suborbits in $X_{\cC}$ under the action of the
    symmetric group $S_r$ {\sl (this enumerates all parity-check
      matrices without redundant rows, up to equivalence)}.
    
  \item For each representative $\bH$ of the suborbits, collect all
    matrices enlarged by adding $\rho-r$ different redundant rows that
    are $\ff_2$-sums of at least two rows of $\bH$.  Let
    $X_{\cC,\rho}$ be the union of all such $\rho\times n$ matrices.

  \item Determine the orbits in $X_{\cC,\rho}$ under the action of the
    symmetric group $S_{\rho}$, and output a representative for each
    orbit.
  \end{enumerate}
\end{enumerate}

\noindent\hrulefill\smallskip

\end{algorithm}

This algorithm was implemented in the C programming language.  The
minimum pseudoweights for the various parity-check matrices were
computed by using Maple 12 and the Convex package~\cite{maple-convex}.

\subsection{Results}\label{sec:short_codes_results}

\begin{table}
  \caption{The Number of Binary $[n,k,d]$ Codes
    with~$d\ge 3$~and~without~Zero~Coordinates}
  \label{table:no-codes}
  {\centering
    \begin{tabular}{r|rcccc}
      & $k=1$ & $2$ & $3$ & $4$ & $5$ \\\hline
      & & & & & \vspace{-2mm} \\
      $n=5$ & $1$ & $1$ & & & \\
      $6$ & $1$ & $3$ & $1$ & & \\
      $7$ & $1$ & $4$ & $4$ & $1$ & \\
      $8$ & $1$ & $6$ & $10$ & $5$ & \\
      $9$ & $1$ & $8$ & $23$ & $23$ & $5$
    \end{tabular}\\}%
\end{table}

We considered all binary linear codes up to length~$n$ with minimum
distance $d\ge 3$ and without zero coordinates, up to code
equivalence.  The number of those codes for given length~$n$ and
dimension~$k$ is shown in Table~\ref{table:no-codes}.

\subsubsection{AWGNC pseudoweight}

The following results were found to hold for all codes of length $n\le
9$.

\begin{itemize}\setlength{\parsep}{0pt}\setlength{\itemsep}{4pt}
\item There are only two codes $\cC$ with $\rho_{\AWGNC}(\cC)>r$,
  i.e., in class $0$ or $1$ for the AWGNC.
  \begin{itemize}
  \item The $[8,4,4]$ extended Hamming code is the shortest code
    $\cC$ in class 1.  We have $\rho_{\AWGNC}(\cC)=5>4=r$ and out
    of $12$ possible parity-check matrices (up to equivalence) with one
    redundant row there is exactly one matrix $\bH$ with
    $\weight_{\min}^{\AWGNC}(\bH) = 4$, namely
    \[ \bH = \mat{
      \ze & \zo & \zo & \ze & \ze & \zo & \zo & \ze \\
      \zo & \ze & \zo & \ze & \zo & \ze & \zo & \ze \\
      \zo & \zo & \ze & \ze & \zo & \zo & \ze & \ze \\
      \ze & \ze & \ze & \ze & \zo & \zo & \zo & \zo \\
      \zo & \zo & \zo & \zo & \ze & \ze & \ze & \ze \\
    }\:. \] There is exactly one matrix $\bH$ with
    $\weight_{\min}^{\AWGNC}(\bH) = 25/7$, and for the remaining
    matrices $\bH$ we have $\weight_{\min}^{\AWGNC}(\bH) = 3$.

    For this code, also $\rho_{\BEC}(\cC)=5>4$, and it is the only
    code of length $n\le 9$ with $\rho_{\BEC}(\cC)>r$.
  \item Out of the four $[9,4,4]$ codes there is one code $\cC$ in
    class 1.  We have $\rho_{\AWGNC}(\cC)=6>5=r$ and out of $2526$
    possible parity-check matrices (up to equivalence) with one
    redundant row there are $13$ matrices $\bH$ with
    $\weight_{\min}^{\AWGNC}(\bH) = 4$.
  \end{itemize}
\item For all codes $\cC$ of minimum distance $d\ge 3$ and for all
  parity-check matrices $\bH$ of $\cC$ we have
  $\weight_{\min}^{\AWGNC}(\bH)\ge 3$; in particular, if $d=3$,
  then $\cC$ is in class~$3$ for the AWGNC.
\item For the $[7,3,4]$ simplex code there is (up to equivalence) only
  one parity-check matrix $\bH$ without redundant rows such that
  $\weight_{\min}^{\AWGNC}(\bH) = 4$, namely
  \[ \bH = \mat{
    \ze & \ze & \zo & \ze & \zo & \zo & \zo \\
    \zo & \ze & \ze & \zo & \ze & \zo & \zo \\
    \zo & \zo & \ze & \ze & \zo & \ze & \zo \\
    \zo & \zo & \zo & \ze & \ze & \zo & \ze \\
  }\:. \] It is the only parity-check matrix with constant row weight
  $3$.
\end{itemize}

\subsubsection{BSC pseudoweight}

We computed the pseudocodeword redundancy for the BSC for all codes of
length $n\le 8$.

\begin{itemize}\setlength{\parsep}{0pt}\setlength{\itemsep}{4pt}
\item The shortest codes with $\rho_{\BSC}(\cC)>r$, i.e., in class~$0$
  or $1$ for the BSC, are the $[7,4,3]$ Hamming code $\cC$ and its
  dual code $\cC^{\bot}$, the $[7,3,4]$ simplex code.  We have
  $\rho_{\BSC}(\cC)=4>3$ and $\rho_{\BSC}(\cC^{\bot})=5>4$.
\item There are two codes of length~$8$ with
  $\rho_{\BSC}(\cC)>r$.  These are the $[8,4,4]$ extended Hamming
  code, for which $\rho_{\BSC}(\cC)=6>4$ holds, and one of the three
  $[8,3,4]$ codes, which satisfies $\rho_{\BSC}(\cC)=6>5$.
\end{itemize}

\subsubsection{Max-fractional weight}

We computed the pseudocodeword redundancy with respect to the
max-fractional weight for all codes of length $n\le 8$.

\begin{itemize}\setlength{\parsep}{0pt}\setlength{\itemsep}{4pt}
\item The shortest code with $\rho_{\maxfrac}(\cC)>r$ is the unique
  $[6,3,3]$ code $\cC$.  We have $\rho_{\maxfrac}(\cC)=4>3$.
\item There are two codes of length~$7$ with $\rho_{\maxfrac}(\cC)>r$.
  These are the $[7,4,3]$ Hamming code and the $[7,3,4]$ simplex code,
  which have both pseudocodeword redundancy~$7$.  In both cases, there
  is, up to equivalence, a unique parity-check matrix $\bH$ with seven
  rows that satisfies $d(\cC)=w_{\min}^{\maxfrac}(\bH)$.  

  This demonstrates that Propositions~\ref{prop:Hamming}
  and~\ref{prop:simplex} are sharp for the max-fractional weight, and
  that the parity-check matrices constructed in the proofs are unique
  in this case.
\item For the $[8,4,4]$ extended Hamming code $\cC$ we have
  $\rho_{\maxfrac}(C)=\infty$, and thus the code is in class~$0$ for
  the max-fractional weight.  It is the shortest code with infinite
  max-fractional pseudoredundancy.

  (It can be checked that $\bldx=[1,1,1,1,1,1,1,3]$ is a
  pseudocodeword in $\cK(\bH)$, where the rows of $\bH$ consist of all
  dual codewords; since $\weight_{\maxfrac}(\bldx) = \frac{10}{3}
  < 4$, we have $\weight_{\min}^{\maxfrac}(\bH) < 4$.)
\item There are two other codes of length~$8$ with
  $\rho_{\maxfrac}(\cC)>r$, namely two of the three $[8,3,4]$ codes,
  having pseudocodeword redundancy~$6$ and~$8$, respectively.
\end{itemize}

\subsubsection{Comparison}

Comparing the results for the AWGNC and BSC pseudoweights, and the
max-fractional weight, we can summarize the results as follows.

\begin{itemize}\setlength{\parsep}{0pt}\setlength{\itemsep}{4pt}
\item For the $[7,4,3]$ Hamming code $\cC$ we have
  $\rho_{\AWGNC}(\cC)=r=3$\,, $\rho_{\BSC}(\cC)=4$\,, and
  $\rho_{\maxfrac}(\cC)=7$.
\item For the $[7,3,4]$ simplex code $\cC$ we have
  $\rho_{\AWGNC}(\cC)={r=4}$\,, $\rho_{\BSC}(\cC)=5$\,, and
  $\rho_{\maxfrac}(\cC)=7$.  
\item For the $[8,4,4]$ extended Hamming code $\cC$ we have
  $\rho_{\AWGNC}(\cC)=5$\,, $\rho_{\BSC}(\cC)=6$\,, and
  $\rho_{\maxfrac}(\cC)=\infty$.  This code $\cC$ is the shortest
  one such that $\rho_{\AWGNC}(\cC)>r$, and also the shortest one
  such that $\rho_{\maxfrac}(\cC)=\infty$.
\item If $d\ge 3$ then for \emph{every} parity-check matrix $\bH$ we
  have $\weight_{\min}^{\AWGNC}(\bH)\ge 3$.  This is not true for
  the BSC and the max-fractional weight.
\end{itemize}

These observations show that there is some significant difference
between the various types of pseudocodeword redundancies.

It is also interesting to note that the $[7,4,3]$ Hamming code is
geometrically perfect, while the $[7,3,4]$ code and the $[8,4,4]$ code
are not (cf.\ Section~\ref{sec:related}).



\section{Cyclic Codes Meeting the Eigenvalue Bound}%
\label{sec:kv_bound}

In this last section we apply the eigenvalue-based lower bound on the
minimum AWGNC pseudoweight by Vontobel and Koetter
\cite{KV-lower-bounds}, see Proposition~\ref{prop:KV_bound}.  We
investigate for which cyclic codes of short length this bound is sharp
with respect to the minimum Hamming distance, for in this case, the
codes have finite AWGNC pseudoredundancy.

We consider binary cyclic codes with full circulant parity-check
matrices, defined as follows: Let $\cC$ be a binary cyclic code of
length $n$ with \emph{check polynomial} $h(x)=\sum_{i\in\cI}h_ix^i$
(cf.\ \cite{MacWilliams_Sloane}, p.~194). Then the \emph{full
  circulant parity-check matrix} for~$\cC$ is the $n\times n$ matrix
$\bH=(H_{j,i})_{i,j\in \cI}$ with entries $H_{j,i}=h_{j-i}$.  Here,
all the indices are modulo $n$, so that $\cI=\{0,1,\dots,n-1\}$.

Since such a matrix is $w$-regular, where $w=\sum_{i\in\cI}h_i$, we
may use the eigenvalue-based lower bound of
Proposition~\ref{prop:KV_bound} to examine the AWGNC pseudocodeword
redundancy: If the right hand side equals the minimum distance $d$ of
the code~$\cC$, then $\rho_{\AWGNC}(\cC)\le n$.

Note that the largest eigenvalue of the matrix $\bL = \bH^T \bH$ is
$\mu_1=w^2$, since every row weight of $\bL$ equals
$\sum_{i,j\in\cI}h_ih_j=w^2$.  Consequently, the eigenvalue bound is
\[ \weight_{\min}^{\AWGNC} \ge n \cdot \frac{2w-\mu_2}{w^2-\mu_2}
\:, \] where $\mu_2$ is the second largest eigenvalue of $\bL$.  We
remark further that $\bL=(L_{j,i})_{i,j\in\cI}$ is a symmetric
circulant matrix, with $L_{j,i}=\ell_{j-i}$ and
$\ell_i=\sum_{k\in\cI}h_kh_{k+i}$.  The eigenvalues of $\bL$ are thus
given by \[ \lambda_j = \sum_{i\in\cI}\ell_i\zeta_n^{ij} =
\operatorname{Re}\sum_{i\in\cI}\ell_i\zeta_n^{ij} =
\sum_{i\in\cI}\ell_i\cos(2\pi ij/n) \] for $j\in\cI$, where
$\zeta_n=\exp(2\pi{\bf i}/n)$ is the $n$-th primitive root of unity
and ${\bf i}^2=-1$ (see, e.g., \cite{Davis-book}, Theorem~3.2.2).

We also consider quasi-cyclic codes of the form given in the following
remark.  This code construction is only introduced for completeness
towards classifying the results; the resulting codes are not
interesting for applications, as the minimum Hamming distance is at
most~$2$ for $m\ge 2$.

\begin{remark}\label{rem:qc}
  Denote by $\bldone_m$ the $m\times m$ matrix with all entries equal to
  $1$.  If $\bH$ is a $w$-regular circulant $n\times n$ matrix then
  the Kronecker product $\tilde{\bH} \define \bH\otimes\bldone_m$ will
  be a $w$-regular circulant $mn\times mn$-matrix and defines a
  quasi-cyclic code.  We have
  \[ \tilde{\bL} = \tilde{\bH}^T\tilde{\bH} = \bH^T\bH \otimes
  \bldone_m^T\bldone_m = \bL \otimes (m\bldone_m)\:,\] and the eigenvalues
  of $m\bldone_m$ are $m^2$ and $0$.  Thus, the largest eigenvalues of
  $\tilde{\bL}$ are $\tilde{\mu}_1 = m^2\mu_1 = m^2w^2$ and
  $\tilde{\mu}_2 = m^2\mu_2$, and the eigenvalue bound of
  Proposition~\ref{prop:KV_bound} becomes
  \[ \weight_{\min}^{\AWGNC} \ge mn \cdot 
  \frac{2mw-m^2\mu_2}{m^2w^2-m^2\mu_2} = 
  n \cdot \frac{2w-m\mu_2}{w^2-\mu_2} \:. \]
\end{remark}

We carried out an exhaustive search on all cyclic codes $\cC$ up to
length $n\le 250$ and computed the eigenvalue bound in all
cases where the Tanner graph of the full circulant parity-check matrix
is connected, by using the following algorithm.\smallskip

\begin{algorithm}

\hrulefill\smallskip

\textbf{Input:} Parameter $n$ (code length).

\textbf{Output:} For all divisors of $x^n-1$, corresponding to cyclic
codes $\cC$ with full circulant parity-check matrix, such that the
Tanner graph is connected: the value of the eigenvalue bound.\medskip

\begin{enumerate}
\item Factor $x^n-1$ over~$\ff_2$ into irreducibles, using Cantor and
  Zassenhaus' algorithm (cf.\ \cite{vzGG-book}, Section~14.3).
\item For each divisor $f(x)$ of $x^n-1$:
\begin{enumerate}
\item Let $f(x)=\sum_ih_ix^i$ and $\bH=(h_{j-i})_{i,j\in\cI}$.
\item Check that the corresponding Tanner graph is connected (i.e.,
  that the greatest common divisor of the indices $i$ with $h_i=1$
  together with $n$ is~$1$).
\item Compute the eigenvalues of~$\bL=\bH^T\bH$: Let
  $\ell_i=\sum_{k\in\cI}h_kh_{k+i}$ and for $j\in\cI$ compute
  $\sum_i\ell_i\cos(2\pi ij/n)$.
\item Determine the second largest eigenvalue $\mu_2$ and output
  $n\cdot(2\ell_0-\mu_2)/(\ell_0^2-\mu_2)$.
\end{enumerate}
\end{enumerate}

\noindent\hrulefill\smallskip

\end{algorithm}

This algorithm was implemented in the C programming language.
Tables~\ref{table:KV_bound-2} and~\ref{table:KV_bound-1} give a
complete list of all cases in which the eigenvalue bound equals the
minimum Hamming distance $d$, for the cases $d=2$ and $d\ge 3$,
respectively.  In particular, the AWGNC pseudoweight equals the
minimum Hamming distance in these cases and thus we have for the
pseudocodeword redundancy $\rho_{\AWGNC}(\cC)\le n$.  All examples of
minimum distance $2$ are actually quasi-cyclic codes as in
Remark~\ref{rem:qc} with parity-check matrix $\tilde{\bH} =
\bH\otimes\bldone_2$.  We list here the constituent code given by the
parity-check matrix~$\bH$.

\begin{table}
  \caption{Binary Cyclic Codes up to Length 250 
    with $d=2$ Meeting~the~Eigenvalue~Bound}\label{table:KV_bound-2}
  {\centering
    \begin{tabular}{ccl}
      parameters & $w$-regular & constituent code\vspace{.5mm} \\\hline
      & & \vspace{-2mm} \\
      $[2n,2n\!-\!m,2]$ & $2^m$ & Hamming c., $n=2^m\!-\!1$, $m=2\dots 6$ \\
      \!\!\!$[2n,2n\!-\!m\!-\!1,2]$\!\!\! & 
      \!\!\!$2^m\!-\!2$\!\!\! & Hamming c. with overall 
      parity-check \\
      $[42,32,2]$ & $10$ & projective geometry code $PG(2,4)$ \\
      $[146,118,2]$ & $18$ & projective geometry code $PG(2,8)$ \\
      $[170,153,2]$ & $42$ & a certain $[85,68,6]$ $21$-regular code \\
      & & (the eigenvalue bound is 5.2) \\
    \end{tabular}\\}%
\end{table}

\begin{table}
  \caption{Binary Cyclic Codes up to Length 250 
    with $d\ge 3$ Meeting~the~Eigenvalue~Bound}\label{table:KV_bound-1}
  {\centering
    \begin{tabular}{ccl}
      parameters & $w$-regular & comments\vspace{.5mm} \\\hline
      & & \vspace{-2mm} \\
      $[n,1,n]$ & $2$ & repetition code, $n=3\dots 250$ \\
      $[n,n\!-\!m,3]$ & \!\!$2^{m-1}$\!\! 
      & Hamming c., $n=2^m\!-\!1$, $m=3\dots 7$ \\
      $[7,3,4]$ & $3$ & dual of the $[7,4,3]$ Hamming code \\
      $[15,7,5]$ & $4$ & Euclidean geometry code EG(2,4) \\
      $[21,11,6]$ & $5$ & projective geometry code PG(2,4) \\
      $[63,37,9]$ & $8$ & Euclidean geometry code EG(2,8) \\
      $[73,45,10]$ & $9$ & projective geometry code PG(2,8) \\
    \end{tabular}\\}%
\end{table}

We conclude this section by proving a result which was observed by the
experiments.

\begin{lemma}
  Let $m\ge 3$ and let $\cC$ be the intersection of a Hamming code of
  length $n=2^m-1$ with a simple parity-check code of length $n$,
  which is a cyclic $[n\,, {n-m-1}\,, 4]$ code.  Consider its full circulant
  parity-check matrix $\bH$.  Then \[ \weight_{\min}^{\AWGNC}(\bH)\ge
  3+\frac{1}{2^{m-2}-1}>3 \:. \]

  In particular, if $m=3$ then $\cC$ is the $[7,3,4]$ code and the
  result implies $\weight_{\min}^{\AWGNC}(\bH)=4$ and $\rho_{\AWGNC}
  (\cC) \le 7$.
\end{lemma}

\addtolength{\textheight}{-8cm}   

\begin{IEEEproof}
  Let $\bH$ be the $w$-regular full circulant parity-check matrix for
  $\cC$.  We claim that $w=2^{m-1}\!-\!1$.  Indeed, each row $\bldh$
  of $\bH$ is a codeword of the dual code $\cC^{\bot}$, and since
  $\cC^{\bot}$ consists of the codewords of the simplex code and their
  complements, the weight of $\bldh$ and thus $w$ must be
  $2^{m-1}\!-\!1$, $2^{m-1}$, or $2^m\!-\!1$.  But $w$ cannot be even,
  for otherwise all codewords of $\cC^{\bot}$ would be of even weight.
  As $w=2^m\!-\!1$ is clearly impossible, it must hold
  $w=2^{m-1}\!-\!1$.

  Next, we show that the second largest eigenvalue of
  $\bL=\bH^T\bH=(L_{j,i})_{i,j\in\cI}$ equals $\mu_2=2^{m-2}$.
  Indeed, let $\bldh_1$ and $\bldh_2$ be different rows of $\bH$,
  representing codewords of $\cC^{\bot}$.  As their weight is equal,
  their Hamming distance is even, and thus it must be $2^{m-1}$.
  Hence, the size of the intersection of the supports of $\bldh_1$ and
  $\bldh_2$ is $2^{m-2}\!-\!1$.  This implies that $L_{i,i}=w$ and
  $L_{j,i}=2^{m-2}\!-\!1$, for $i\ne j$.  Consequently, $\bL$ has an
  eigenvalue of multiplicity $n-1$, namely
  $w-(2^{m-2}\!-\!1)=2^{m-2}$, and thus $\mu_2$ must be $2^{m-2}$.

  Finally, we apply Proposition~\ref{prop:KV_bound} to get
  \begin{align*}
    \weight_{\min}^{\AWGNC} &\ge (2^m\!-\!1)\,
    \frac{2\,(2^{m-1}\!-\!1)-2^{m-2}}{(2^{m-1}\!-\!1)^2-2^{m-2}}\\
    &= \frac{(2^m\!-\!1)\,\big(2\,(2^{m-1}\!-\!1)-2^{m-2})}{(2^m\!-\!1\big)\,
      (2^{m-2}\!-\!1)}\\
    &= \frac{3\,(2^{m-2}\!-\!1)+1}{2^{m-2}-1} \:,
  \end{align*}
  which proves the result.
\end{IEEEproof}


\section*{Acknowledgments}

The authors are indebted to the anonymous reviewers for their
worthwhile suggestions and recommendations, which significantly
improved the exposition of the work, and also to Pascal Vontobel for
carefully handling the manuscript.  In addition, they would like to
thank Nigel Boston, Christine Kelley, and Olgica Milenkovic for helpful
discussions.


\end{document}